\documentclass[12pt]{article}
\usepackage{epsfig}
\usepackage{here}
\usepackage{a4}
\usepackage{amssymb}

\begin{document}

\def\Journal#1#2#3#4{{#1} {\bf #2}, #3 (#4)}

\def\etal{{\it et\ al.}}
\def\NCA{\em Nuovo Cim.}
\def\NIM{\em Nucl. Instrum. Methods}
\def\NIMA{{\em Nucl. Instrum. Methods} A}
\def\NPB{{\em Nucl. Phys.} B}
\def\PLB{{\em Phys. Lett.}  B}
\def\PRL{\em Phys. Rev. Lett.}
\def\PRC{{\em Phys. Rev.} C}
\def\PRD{{\em Phys. Rev.} D}
\def\ZPC{{\em Z. Phys.} C}
\def\ASP{{\em Astrop. Phys.}}
\def\JETP{{\em JETP Lett.\ }}

\def\numunue{\nu_\mu\rightarrow\nu_e}
\def\numunutau{\nu_\mu\rightarrow\nu_\tau}
\def\nuebar{\bar\nu_e}
\def\nue{\nu_e}
\def\nutau{\nu_\tau}
\def\numubar{\bar\nu_\mu}
\def\numu{\nu_\mu}
\def\ra{\rightarrow}
\def\numubarnuebar{\bar\nu_\mu\rightarrow\bar\nu_e}
\def\nuebarnumubar{\bar\nu_e\rightarrow\bar\nu_\mu}
\def\osc{\rightsquigarrow}

\def\inteni{{\cal I}_{pot}}
\def\fmerit{{\cal F}}

\thispagestyle{empty}
\begin{flushright}
{\tt ICARUS/TM-2002/10}\\ 
\today 
\end{flushright}
\vspace*{0.5cm}
\begin{center}
{\Large{\bf Proton driver optimization for 
new generation neutrino superbeams
to search for sub-leading
$\numu\ra\nue$ oscillations ($\theta_{13}$ angle)} }\\
\vspace{.5cm}
A. Ferrari$^{a,}$\footnote{On leave of absence from INFN Milano.},
A. Rubbia$^b$,
C. Rubbia$^c$
and P.R. Sala$^{b,1}$

\vspace*{0.3cm}
a) CERN, Geneva,
Switzerland \\
b) Institut f\"{u}r Teilchenphysik, ETHZ, CH-8093 Z\"{u}rich,
Switzerland\\
c) ENEA and Pavia University, Italy
\end{center}
\vspace{0.3cm}
\begin{abstract}
\noindent
In this paper, we 
perform a systematic study of particle production and 
neutrino yields 
for different incident proton energies $E_p$ and baselines $L$, with the aim 
of optimizing  the parameters of a neutrino beam for the investigation of    
$\theta_{13}$-driven neutrino oscillations in  the 
$\Delta m^2$ range allowed by Superkamiokande results. 
We study the neutrino energy spectra in the ``relevant'' region of the
first maximum of the oscillation at a given baseline $L$.
We find that  to each baseline $L$ corresponds an ``optimal''
proton energy $E_p$ which minimizes the required integrated
proton intensity needed to observe a fixed number of oscillated
events. 
In addition, we find that the neutrino event rate in the relevant region scales
approximately linearly with the proton energy. 
Hence, baselines $L$ and proton
energies $E_p$ can be adjusted and the performance for neutrino
oscillation searches will remain approximately unchanged provided
that the product of the proton energy times the number of protons on
target remains constant. 
We apply these ideas to the specific cases
of 2.2, 4.4, 20, 50 and 400~GeV protons. We simulate focusing systems
that are designed to best capture the secondary pions of the ``optimal''
energy. We compute the expected sensitivities to $\sin^22\theta_{13}$ 
for the various configurations by assuming the existence
of new generation accelerators able to deliver
integrated proton intensities on target times the proton energy 
of the order of ${\cal O}(5\times 10^{23})\rm\ GeV\times\rm pot/year$.
\end{abstract}

\pagestyle{plain} 
\setcounter{page}{1}
\setcounter{footnote}{0}

\section{Introduction}
The firmly established disappearance of muon neutrinos of cosmic ray 
origin~\cite{kamio,Fukuda:1998mi} strongly points 
toward the existence of neutrino
oscillations~\cite{pontecorvo}. 

The approved first generation long baseline (LBL) experiments ---
K2K~\cite{k2k}, MINOS~\cite{minos}, 
ICARUS~\cite{icarus} and OPERA~\cite{opera} --- will search for
a conclusive and unambiguous signature of the oscillation
mechanism.
They will provide the first precise
measurements of the parameters governing the main
muon disappearance mechanism. 
In particular, the CERN-NGS beam\cite{CNGS,Addendum}, specifically optimized for tau appearance,
will allow to directly confirm the hints for neutrino flavor oscillation.

In addition to the dominant $\numu\ra\nutau$ oscillation, it is possible that a sub-leading
transition involving electron-neutrinos occur as well.
In the ``standard interpretation'' of the 3-neutrino mixing, the $\numu \ra \nue$ oscillations at the 
$\Delta m^2 \approx 2.5\times 10^{-3}\rm\ eV^2$ indicated
by atmospheric neutrinos is driven by the so-called $\theta_{13}$ angle.
Indeed, given
the flavor eigenstates
$\nu_\alpha(\alpha= e,\mu,\tau)$ related to the mass eigenstates
$\nu'_i(i=1,2,3)$ where
$\nu_\alpha=U_{\alpha i}\nu'_i$, the mixing matrix $U$ is parameterized as:
\begin{equation}
U(\theta_{12},\theta_{13},\theta_{23},\delta)=\left(
\begin{tabular}{ccc}
$c_{12}c_{13}$      & $s_{12}c_{13}$   &  $s_{13}e^{-i\delta}$ \\
$-s_{12}c_{23}-c_{12}s_{13}s_{23}e^{i\delta}$ &
$c_{12}c_{23}-s_{12}s_{13}s_{23}e^{i\delta}$ & $c_{13}s_{23}$ \\
$s_{12}s_{23}-c_{12}s_{13}c_{23}e^{i\delta}$ &
$-c_{12}s_{23}-s_{12}s_{13}c_{23}e^{i\delta}$ & $c_{13}c_{23}$ 
\end{tabular}\right)
\end{equation}
with $s_{ij}=\sin\theta_{ij}$ and $c_{ij}=\cos\theta_{ij}$.

The best sensitivity for this oscillation is expected for ICARUS at the CERN-NGS.
Limited by the CNGS beam statistics at low energy, this search
should allow to improve by roughly a factor 5 (see Ref.\cite{icarus}) 
the CHOOZ\cite{Chooz}
limit on the $\theta_{13}$ angle for $\Delta m^2 \approx 3\times 10^{-3}\rm\ eV^2$.
Beyond this program, new methods will be required in order to improve significantly
the sensitivity. 

At present, the only well established proposal in this direction 
is the JHF-Kamioka project\cite{JHF}. 
In its first phase, 5 years of operation with
the Super-K detector, it aims to a factor 20 improvement
over the  CHOOZ limit.

In Ref.\cite{Rubbia:2002rb}, we have studied an optimization of the CNGS optics that would
allow to increase the neutrino flux yield at low energy by a factor 5 compared
to the baseline $\tau$-optimization of the CNGS beam. This would yield an improvement
in the sensitivity by about a factor two, or equivalently an improvement of a factor
10 compared to CHOOZ.

In this paper, FLUKA\cite{FLUKA1,FLUKA2} Monte Carlo 
simulations are employed 
to perform a systematic study of particle production and 
neutrino yields 
for different beam energy and baselines.   
Focusing systems adapted to the
low energy range are also investigated, to obtain realistic estimates of
the achievable neutrino rates.    
 
The whole procedure is assumed to be detector and accelerator independent. 
Nevertheless, 
to give  a first estimate of the needed beam intensities, the ICARUS T3000
detector size has been assumed as a reference. This is also justified by
a recent neutrino detector comparison\cite{DHarris}, following which 
the  2.35~kton LAr ICARUS fiducial mass is equivalent to  a detector of 
approximately 20~kton of steel or 50~kton of water.  For this reason, we concentrate
on the intrinsic electron-neutrino background from the beam and do
not explicitly calculate other sources of backgrounds that are strongly related to detector
performances, such as backgrounds from
 $\pi^0$ production in neutral current events . We assume that in reality
the ``best'' beam would be complemented
by the ``best'' detector and that these backgrounds will only introduce
a small correction to our sensitivity estimates. In addition, we are primarily
interested at this stage in the comparison among proton drivers.

We consider so-called ``conventional'' neutrino superbeams, in
which neutrinos are produced by the decay of secondary pions obtained in
high-energy collisions of protons on an appropriate target and followed by
a magnetic focusing system.

In this kind of beams, the neutrino beam spectrum and its flux are essentially
determined by three parameters that can be optimized appropriately:
\begin{itemize}
\item the primary proton energy $E_p$,
\item the number of protons on target $N_{pot}$ per year,
\item the focusing system, which focus a fraction of the secondary charged
pions and kaons (positive or both signs depending on the focusing device).
\end{itemize}

In order to simplify the problem, we consider three ``classes'' of proton energies:
\begin{itemize}
\item {\it Low energy}: protons in a range of a few GeV. We take as reference
the CERN-SPL proton driver design\cite{SPL} with 2.2~GeV kinetic energy
and an ``upgraded SPL'' with similar characteristics but with 4.4~GeV protons;
\item {\it Medium energy}: we take 20~GeV proton energy, similar to the CERN PS machine
and the 50~GeV of the JHF facility\cite{JHF};
\item {\it High-energy}: we take the highest energies, i.e. the
400 GeV protons like in the case of the CERN SPS.
\end{itemize}

The purpose of this work is to understand the required intensities
for the various proton energies, in other words, is it more favorable
to employ low-energy, high-intensity proton or high-energy, low-intensity proton machines?


\section{Observing the oscillation -- choosing $L$ and $E$}
In order to maximize the probability of an oscillation, we must choose the energy of the neutrino $E_{max}$ and
the baseline $L$ such that 
\begin{equation}
1.27\frac{L(km)}{E_{max}(GeV)}\Delta m^2(eV^2)\simeq \frac{\pi}{2}
\end{equation}
However, in order to {\it observe
the oscillation}, we must at least see the maximum preceded by a minimum, given by 
\begin{equation}
1.27\frac{L(km)}{E_{min}(GeV)}\Delta m^2(eV^2)\simeq \pi
\end{equation}

The mass difference indicated by Superkamiokande is given by $\Delta m^2 \approx 2.5\times 10^{-3}\rm\ eV^2$
and lies within the range 
$ 1\times 10^{-3}<\Delta m^2 < 4\times 10^{-3}\rm\ eV^2$\cite{nu2000}. 

Since we are considering one single detector at a given location
as target, our baseline $L$ for a given source
is fixed. This implies that 
{\it the neutrino beam spectrum should not be too narrow}, e.g. it should
be a relatively wide-band beam in order to cope with the uncertainty on the 
$\Delta m^2$. Since the knowledge on this parameter will improve with time, in particular with 
the running of K2K, MINOS and CNGS experiments, the range of energies will eventually
be limited by the desire to observe the minimum and the maximum. In the meantime, we
consider the full mass range indicated by atmospheric neutrino observation.
\par

We do not wish to consider
baselines longer than the current CNGS baseline of $L=730\ \rm km$ and 
use therefore a range of baselines
between 100~km and 730~km.
Table~\ref{tab:enenu}
shows the values (in MeV) of $E_{min}$ and $E_{max}$ defined above as a
function for the baselines. The energies range from about 50 MeV to 300~MeV for $L=100\rm\ km$ and
from 300~MeV to 2400~MeV for $L=730\rm\ km$. Clearly, we wish to optimize for high intensity,
low energy neutrino superbeams. 

We note that at these low energies, neutrino interactions are clearly
identifiable and have generally easily reconstructible final-states. This is an advantage
for detector-related background suppression. In a detector with the granularity
like ICARUS, neutral pions can be easily suppressed via energy ionization or
by direct reconstruction of the two (well separated) decay photons.

\begin{table}[htb]
\begin{center}
\begin{tabular}{|c||c|c||c|c||c|c||c|c|}
\hline
&\multicolumn{8}{c|}{$\Delta m ^2 $(eV $^2$)}\\
\hline
L&\multicolumn{2}{c|}{$1\times 10^{-3}$}&\multicolumn{2}{c|}{$2\times 10^{-3}$}&
\multicolumn{2}{c|}{$3\times 10^{-3}$}&\multicolumn{2}{c|}{$4\times 10^{-3}$}\\
(km)&E$_{max}$  & E$_{min}$  &E$_{max}$  & E$_{min}$  
&E$_{max}$  & E$_{min}$  &E$_{max}$  & E$_{min}$  \\
&MeV&MeV&MeV&MeV&MeV&MeV&MeV&MeV\\
\hline
 100  &    81  &    40  &   162  &    81  &
   243  &   121  &   323  &   162\\
 150  &   121  &    61  &   243  &   121  & 
  364  &   182  &   485  &   243\\
 200  &   162  &    81  &   323  &   162  &
   485  &   243  &   647  &   323\\
 300  &   243  &   121  &   485  &   243  &
   728  &   364  &   970  &   485\\
 400  &   323  &   162  &   647  &   323  &
   970  &   485  &  1294  &   647\\
 500  &   404  &   202  &   809  &   404  &
  1213  &   606  &  1617  &   809\\
 600  &   485  &   243  &   970  &   485  &
  1455  &   728  &  1940  &   970\\
 730  &   590  &   295  &  1180  &   590  &
  1771  &   885  &  2361  &  1180\\
\hline
\end{tabular}
\end{center}
\caption{Neutrino energies $E_{max}$ and $E_{min}$ (see text for definition) 
corresponding to
the maximum and minimum of the $\nu_\mu\rightarrow\nu_x$ oscillation
for various baselines and the $\Delta m^2$ range indicated by
Superkamiokande.}
\label{tab:enenu}
\end{table}


\section{The approximate scaling with proton energy}
The understanding of the relationship between primary beam energy and 
neutrino production is the first step toward any optimization.  To isolate
this relationship from the other parameters, it is more convenient to work
in the ``perfect focusing'' approximation, where {\it all (positively
charged) mesons produced within a given solid angle} 
are supposed to be focused exactly on the
detector direction\footnote{In order to derive the scaling, we consider
only neutrino fluxes. However, in Section~\ref{sec:bngs}, when we 
compute our oscillation sensitivities we shall include both
neutrinos and antineutrinos components obtained from the full
simulation of the mesons of the relevant charge.}. In our case, we set $P_T=0$ for all secondary
positively charged particles produced within 1 rad.

All the simulations presented in this section have been
performed with the same target and decay tunnel geometry. The target is in
graphite, density 1.8 g/cm$^3$, one
meter long, 2~mm radius; the decay tunnel is 150~m long and 3.5~m in radius.
Target optimization will not change the general scaling, and the 
effect of the tunnel length on muon neutrino
production can be accounted
for in first approximation assuming that all neutrinos are produced by pions,
and that they carry the maximum possible momentum. 

\begin{figure}[p]
\centering
\epsfig{file=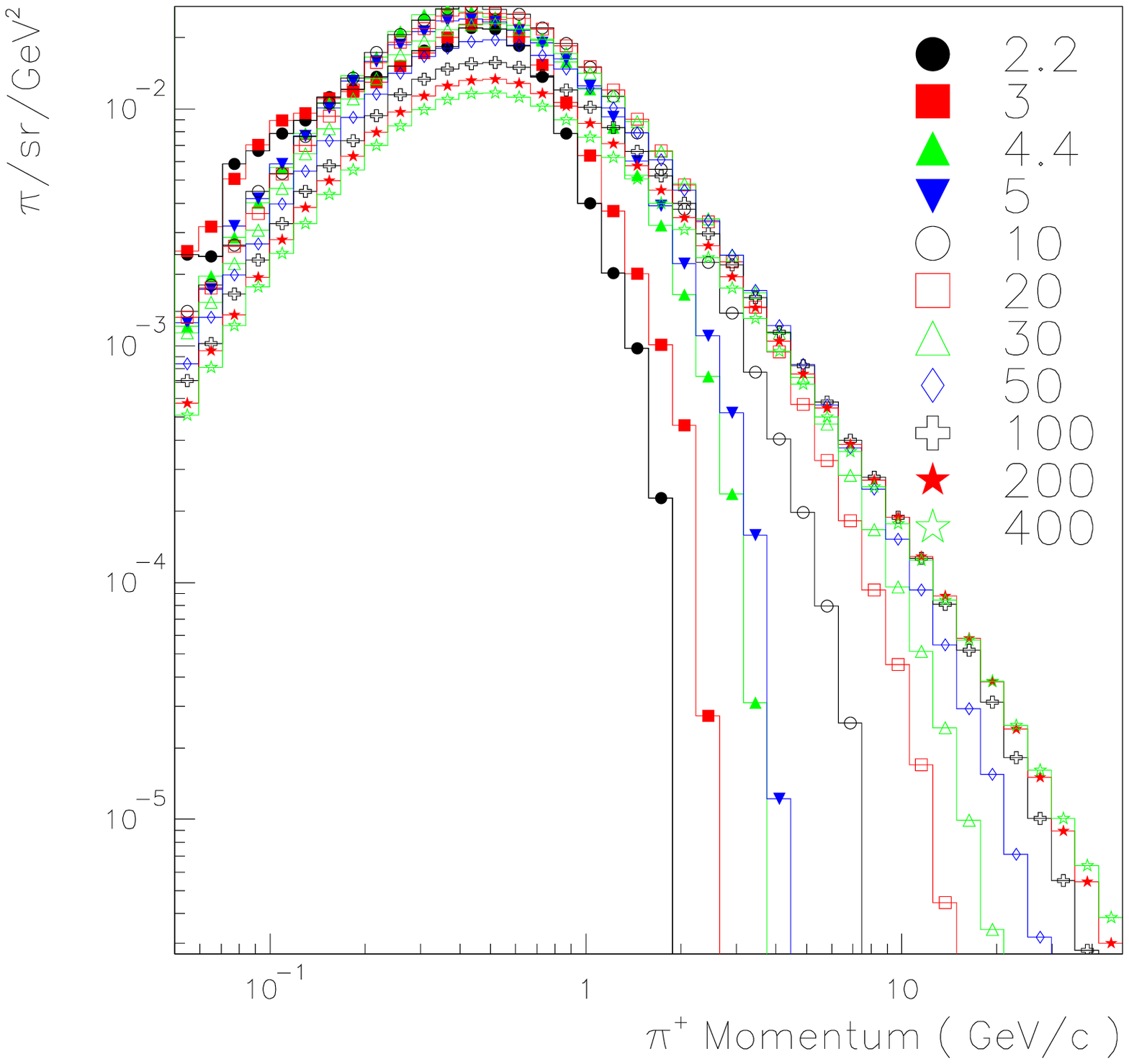,width=15cm}
\caption{Normalized pion production rate $Y_\pi/E_p$ for various
incident proton energies $E_p$ as estimated with FLUKA.}
\label{fig:prodpl}
\end{figure}

The pion production rate at the exit of the target for various
incident proton energies $E_p$ as estimated with FLUKA
are shown in Figure~\ref{fig:prodpl}
for energies ranging from 2.2 GeV up to the 400 GeV of the 
SPS for the CNGS beam. In order to compare pion productions at
different proton energies, we {\it divide the spectra by the proton
energy $E_p$}. 

All normalized spectra have similar  shape, with the maximum
yield at low energies ($p_\pi \approx 500 $~MeV). Far from the endpoint, all
spectra scale approximately with the incoming proton energy. Departures
 from the overall scaling consist in a slightly different shape  at  low
 $E_p$, and harder spectra at  high $E_p$.     

If the detector is far away in the forward direction, the neutrino event
rate as a function of neutrino energy can be derived from these spectra by
considering that:
\begin{itemize}
\item The neutrino carries a momentum that is 0.43 times the parent pion momentum
\item The Lorentz boost gives a factor proportional to $E_\pi^2\propto E_\nu^2$ on the
solid angle   
\item The neutrino cross section grows approximately as $E_\nu$ (not true
at the lowest energies, where the quasi-elastic grows more rapidly, however,
for the optimization of the proton energy, this approximation is adequate.)
\end{itemize}

\begin{figure}[p]
\centering
\epsfig{file=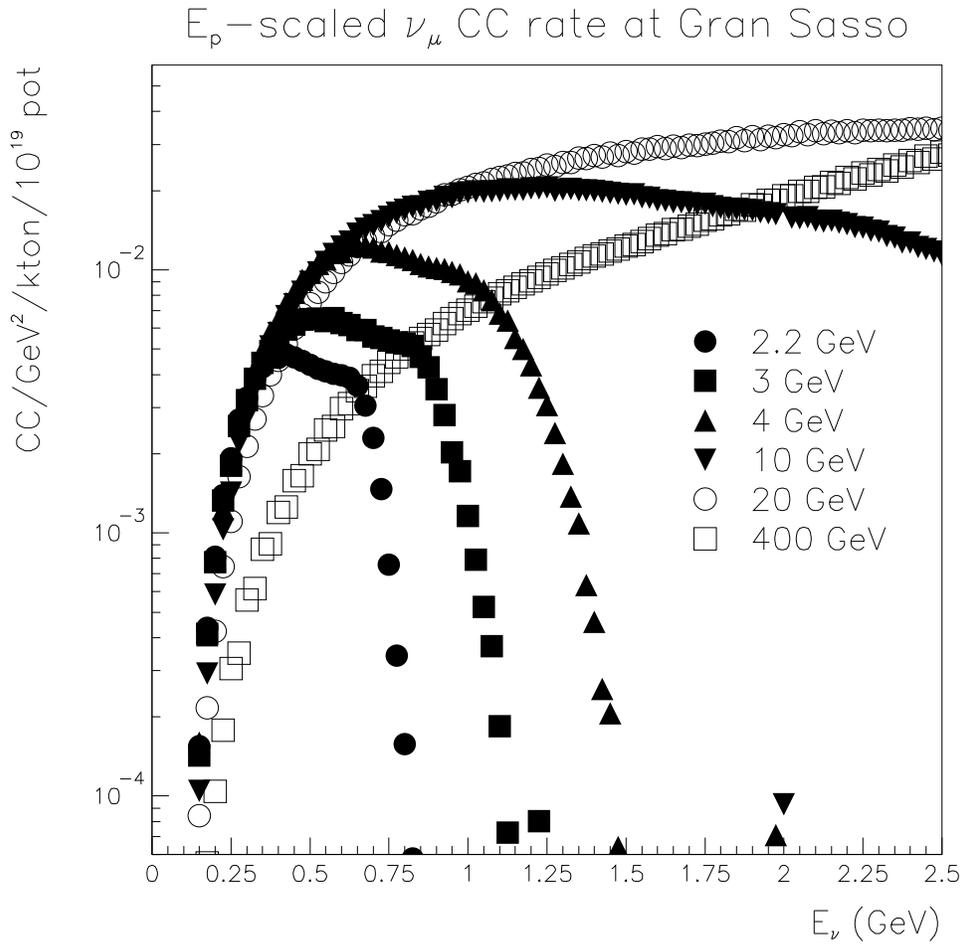,width=15cm}
\caption{Rescaled energy spectrum of charged current events $N_{\numu,CC}(E_\nu)/E_p$
for various incident proton energies $E_{p}$ (arbitrarly normalized to a baseline
$L=732\rm\ km$).
}
\label{fig:splcompr}
\end{figure}

For a given E$_\nu$, these factors apply {\it independently} on the primary
proton energy. Thus, we expect that the energy scaling observed on the
pion production is reflected in the neutrino production. This is 
shown in Figure~\ref{fig:splcompr}, where we plot the simulated muon neutrino
event rate scaled with the primary proton energy.

The super-position of the
curves at the lowest energies is impressive, except for one at $E_p=400\ \rm GeV$. 
Of course, by rising $E_p$ the energy-integrated
event rate raises dramatically because the spectra extend to higher
energies, but the event rate for a given $E_\nu < \approx  0.15\times E_p $ is simply 
proportional to $E_p$. 

This approximate scaling implies that we can define a 
{\it power factor} $\fmerit$ for neutrino
production as the product of the proton energy times the beam intensity:
\begin{equation}
\fmerit = E_p\times N_{pot}
\end{equation}

Up to now we did not take into account the neutrino oscillation
probability. If the aim is that of having a neutrino beam centered on the 
oscillation maximum, $E_\nu$ and the baseline $L$ have to be chosen such as 
\begin{equation}
\sin^2\left( 1.27 \Delta m^2  \frac{L}{E_\nu}\right)\approx 1, 
\end{equation}
thus 
\begin{equation}
E_{max}\propto L.
\end{equation}

\begin{figure}[p]
\centering
\epsfig{file=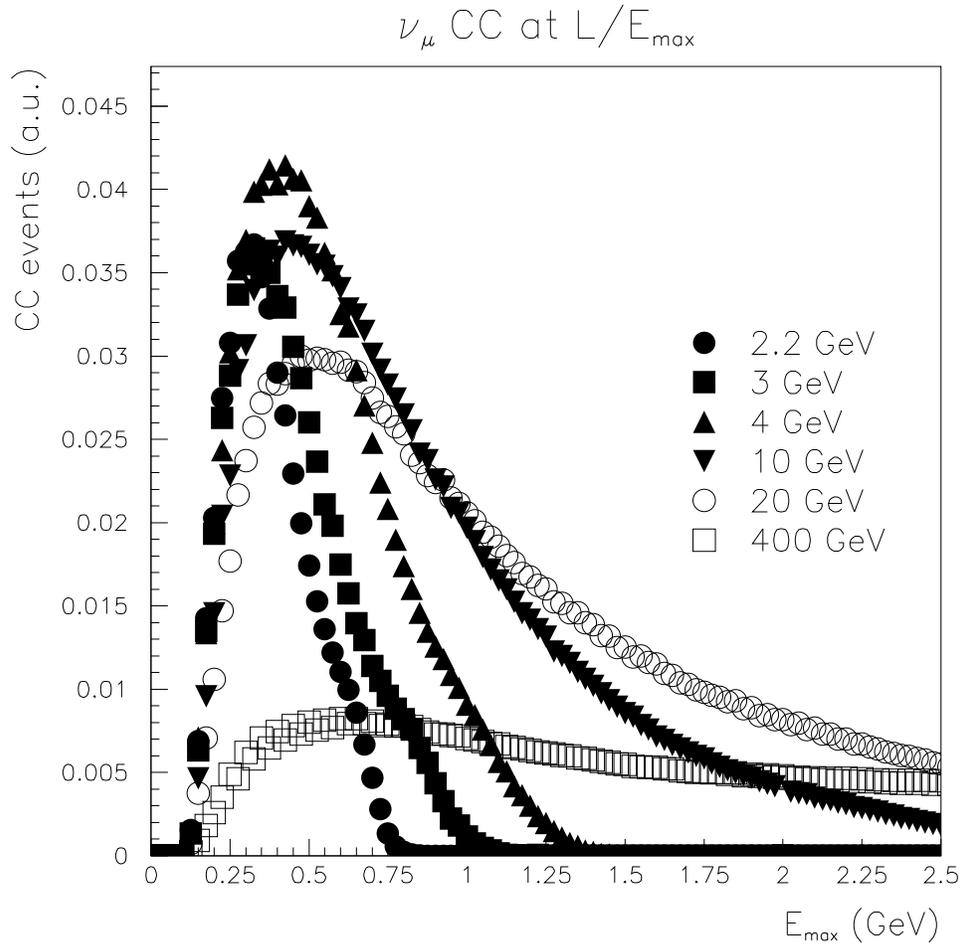,width=15cm}
\caption{Doubly rescaled energy spectrum of charged current 
events $N_{\numu,CC}(E_\nu)/E_\nu^2/E_p$
for various incident proton energies $E_{p}$.
}
\label{fig:splcomplr}
\end{figure}

How to choose the baseline $L$ to maximize the rate of oscillated neutrino events per proton
incident on the neutrino target?  The neutrino flux 
grows scales like $1/L^2$, like the solid angle. 
But $L$ and  $E_{max}$ are proportional, thus  dividing  bin by bin 
 the simulated neutrino spectra  with a factor $E_\nu ^2 $ one gets the 
shape of  the event rate as a function of $E_{max}$. This has been done 
for the perfect focusing approximation in Figure~\ref{fig:splcomplr} for various
proton energies $E_p$. 

It is
evident that the most optimal situation is for $E_{max}$ in the range 200-600~MeV
and {\it this results holds essentially independently of the proton energy}, at least
as long as we discard the very high 400~GeV energy case. We can go beyond
the $1/E_\nu^2$ approximation and in the following we will use the exact expression
for the oscillation probability and will compute exactly the number of oscillated
events by folding the expected neutrino spectra.

Nonetheless, the situation gives us a large freedom in the choice of proton energy, {\it provided
that the intensity can be re-scaled accordingly}, so that the power factor
remains essentially constant
\begin{equation}
\fmerit = E_p\times N_{pot}\approx const.
\end{equation}


\section{The optimal baseline $L$ for each proton energy}
\label{sec:ideafoc}

In order to study the scaling with the proton energy $E_p$ and
the baseline $L$, we compute the number $N_e$ of oscillated $\numu\ra\nue$
events:
\begin{equation}
N_e \propto \inteni \int_0^{\infty} dE_\nu \sigma(E_\nu) \phi(E_\nu) P(\numu\ra\nue,E_\nu,L,\Delta m^2, \sin^22\theta_{13})
\end{equation}
where $\phi(E_\nu)$ is the neutrino flux per proton on target.
We normalize this number to a 2.35 kton (argon) detector, and assume 100\% electron
detection efficiency. For definiteness, we assume unless otherwise noted that
$\Delta m^2=3\times 10^{-3}\rm\ eV^2$ and $\sin^22\theta_{13}=1\times 10^{-3}$.

In order to take into account the effect of focusing (realistic focusing
is discussed in section~\ref{sec:realfoc}), we focus ideally all particles with the
acceptance of 1 rad and
apply a constant ``focusing efficiency'' of 20\%, i.e. scale down the rates
by a factor 5. We will see that this assumption is quite realistic and
at least conservative for the low energy neutrinos.

Similarly, one can also estimate the ``goodness'' of the neutrino
flux by computing the number of muon charged current events
$N^0_{\mu,CC}$
within the energies $E_{min}(\Delta m^2=1\times 10^{-3}\rm\ eV^2)$ and 
$E_{max}(\Delta m^2=4\times 10^{-3}\rm\ eV^2)$ (see Table~\ref{tab:enenu}) of the oscillation:
\begin{equation}
N^0_{\mu,CC} \propto \int_{E_{min}}^{E_{max}} dE_\nu \sigma(E_\nu) \phi(E_\nu)
\end{equation}
We choose to normalize $N^0_{\mu,CC}$ to $10^{19}$ pots and per kton of (argon) target.

With these two parameters, we can easily compare all the possible options.
It is fair to note that the number of oscillated events $N_e$ assumes that
the search is not limited by background, which in practice
is {\it not} the case, as we will show
in section~\ref{sec:bngs}. However, we prefer to separate the issue
of background (which will be discussed in section~\ref{sec:back}) 
and concentrate for the moment on the question of the needed
proton intensity. As we will show, the background turns out to be also quite
independent of the proton energy and hence it does not enter in the 
optimization of the proton energy.

\tabcolsep=.8mm
\begin{table}[tb]
  \begin{center}
    \begin{tabular}{|l|c|c|c|c|c|c|c|c|c|c|}
\hline
\hline
L&\multicolumn{2}{c|}{2.2 GeV}&\multicolumn{2}{c|}{4.4 GeV}
&\multicolumn{2}{c|}{20 GeV}
&\multicolumn{2}{c|}{50 GeV}&\multicolumn{2}{c|}{400 GeV}\\
km& $N^0_{\mu,CC}$ & $\inteni$
& $N^0_{\mu,CC}$ & $\inteni$
& $N^0_{\mu,CC}$ & $\inteni$
& $N^0_{\mu,CC}$ & $\inteni$
& $N^0_{\mu,CC}$ & $\inteni$\\
\hline
 100  &  0.035  & $  2.9\cdot 10^{24}$
  &  0.048  & $  2.1\cdot 10^{24}$
  &   0.20  & $  5.1\cdot 10^{23}$
  &    0.3  & $  3.4\cdot 10^{23}$
  &    1.0  & $  1.0\cdot 10^{23}$
\\
 150  &  0.052  & $  {\bf 1.9\cdot 10^{24}}$
  &  0.098  & $  1.0\cdot 10^{24}$
  &   0.45  & $  2.2\cdot 10^{23}$
  &    0.7  & $  1.4\cdot 10^{23}$
  &    2.3  & $  4.4\cdot 10^{22}$
\\
 200  &  0.046  & $  2.2\cdot 10^{24}$
  &  0.122  & $  {\bf 8.2\cdot 10^{23}}$
  &   0.65  & $  1.5\cdot 10^{23}$
  &    1.1  & $  9.5\cdot 10^{22}$
  &    3.5  & $  2.9\cdot 10^{22}$
\\
 300  &  0.023  & $  4.3\cdot 10^{24}$
  &  0.112  & $  8.9\cdot 10^{23}$
  &   0.90  & $  1.1\cdot 10^{23}$
  &    1.6  & $  6.4\cdot 10^{22}$
  &    5.4  & $  1.9\cdot 10^{22}$
\\
 400  &  0.013  & $  7.7\cdot 10^{24}$
  &  0.081  & $  1.2\cdot 10^{23}$
  &   0.97  & $  {\bf 1.0\cdot 10^{23}}$
  &    1.8  & $  5.6\cdot 10^{22}$
  &    6.4  & $  1.6\cdot 10^{22}$\\
 500  &  0.008  & $  1.2\cdot 10^{25}$
  &  0.055  & $  1.8\cdot 10^{24}$
  &   0.96  & $  {\bf 1.0\cdot 10^{23}}$
  &    1.9  & $  5.3\cdot 10^{22}$
  &    7.1  & $  1.4\cdot 10^{22}$
\\
 600  &  0.006  & $  1.8\cdot 10^{25}$
  &  0.038  & $  2.7\cdot 10^{24}$
  &   0.91  & $  1.1\cdot 10^{23}$
  &    1.9  & $  {\bf 5.2\cdot 10^{22}}$
  &    7.5  & $  {\bf 1.3\cdot 10^{22}}$
\\
 730  &  0.003  & $  2.9\cdot 10^{25}$
  &  0.025  & $  4.0\cdot 10^{24}$
  &   0.83  & $  1.2\cdot 10^{23}$
  &    1.9  & $  {\bf 5.2\cdot 10^{22}}$
  &    7.8  & $  {\bf 1.3\cdot 10^{22}}$
\\
\hline
   \end{tabular}
\caption{Integrated $\nu_\mu $CC events per kton
and $10^{19}$ p.o.t within the relevant energy interval $N^0_{\mu,CC}$  and
integrated beam intensity $\inteni$, assuming
a constant 20\% focusing efficiency wrt perfect focusing, needed to obtain
$N_e=5$ events
(see text) as a function of the baseline $L$ for various
proton energies. 
The configuration corresponding to the minimum
$\inteni$ is shown in bold.}
\label{tab:rateosci}
  \end{center}
\end{table}

The results as a function of the baseline $L$
are summarized in Table~\ref{tab:rateosci}. The configuration corresponding to the minimum
$\inteni$ is shown in bold.

\begin{figure}[p]
\begin{center}
\mbox{\epsfig{file=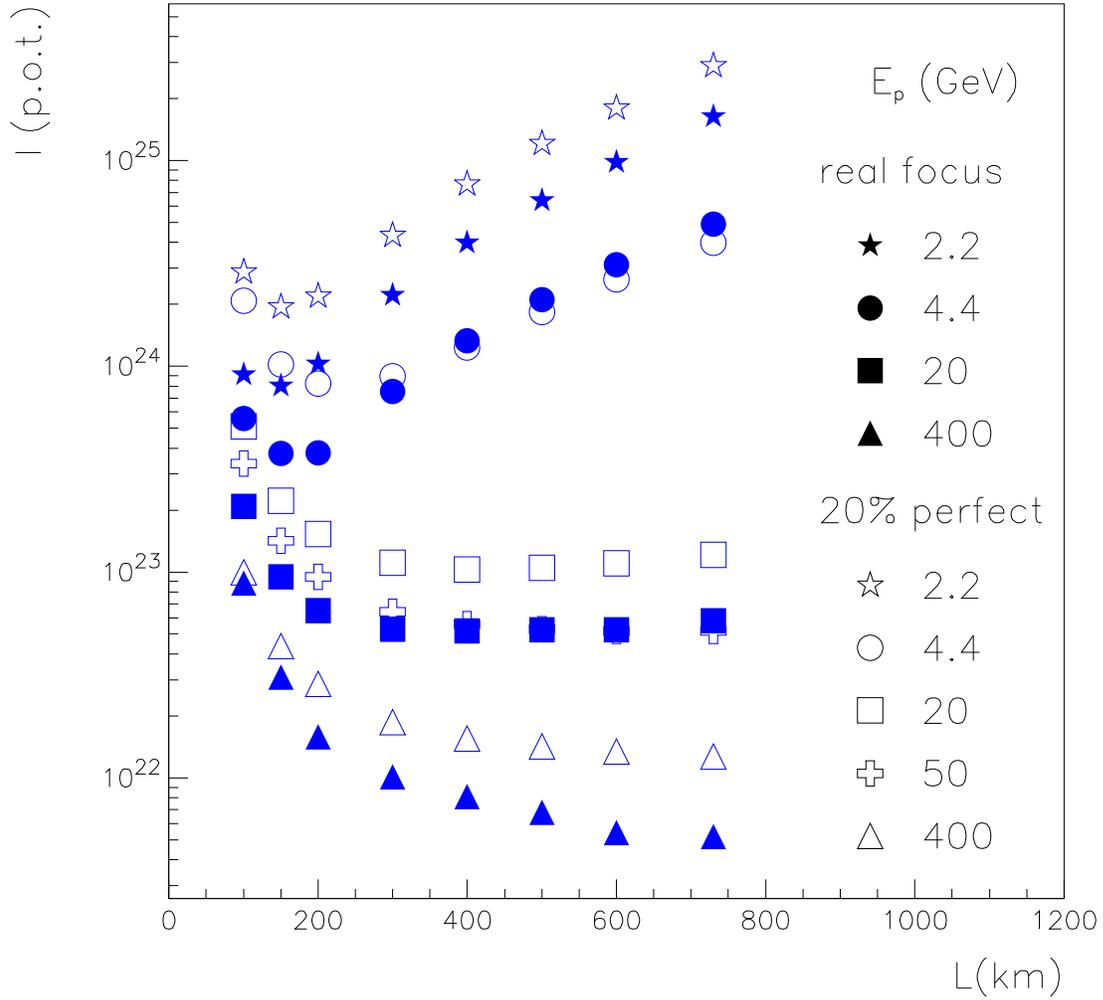,width=17cm}}\end{center}
\caption{ Integrated beam intensity needed to obtain 5 oscillated $\nu_e$ events in a 2.35 kton detector
for  $\Delta m^2=3\times 10^{-3}\rm\ eV^2$ and $\sin^2(2\theta_{13}) = 10^{-3}$ for
different beam energies and baselines. Full symbols correspond to the real
focusing systems described in the text, open symbols to 
energy-independent  20\% scaling from perfect focusing.}
\label{fig:rateosci}
\end{figure}

We observe that {\it for each baseline there is an optimal proton energy $E_p^{optimal}$, which minimizes
the required integrated proton intensity $\inteni$} to observe a fixed number of oscillated events. 
This is also visible in Figure~\ref{fig:rateosci}, where we plot
the integrated beam intensity needed to obtain 5 oscillated $\nu_e$ events in a 2.35 kton detector
for  $\Delta m^2=3\times 10^{-3}\rm\ eV^2$ and $\sin^2(2\theta_{13}) = 10^{-3}$ for
different beam energies and baselines. 
The point of minimum pots corresponds approximately to the maximum $\nu_\mu$ event rate.

This is because in that point, the secondary pion yield 
energy spectrum per proton on target,
is best matched to the neutrino oscillation probability. Conversely, for each proton energy there is
an optimal baseline $L^{opt}$, which maximizes the integrated
neutrino oscillation probability in the neutrino
energy region which corresponds to the largest weighted pion yield at that proton energy.

For a 2.2~GeV proton driver, the optimal baseline $L^{opt}$ is approximately $L^{opt}\approx 150\ \rm km$. 
For a 4.4~GeV proton driver, it is approximately $L^{opt}\approx 200\ \rm km$. 
For 20~GeV, we find $L^{opt}\approx 450\ \rm km$. For energies above 50~GeV, the optimal baseline
is around 700~km. 

For actual baselines smaller than the optimal baseline, $L<L^{opt}$, the neutrino oscillation maximum
occurs at lower energy and the yield of corresponding pions for the given proton energy is lower than
in the optimal case. Hence, we need a higher intensity to compensate for this effect. 

For actual baselines greater than the optimal baseline, $L>L^{opt}$, the neutrino oscillation maximum
occurs at higher energy and the yield of corresponding pions for the given proton energy is lower than
in the optimal case. Indeed, at some point, the optimal neutrino oscillation energy corresponds to
a pion energy, which is kinematically forbidden for the given incident proton energy.
Hence, we need again a higher proton intensity to compensate for the kinematical suppression. 

At the optimal baselines $L_{opt}$ the power factors are the following:
\begin{eqnarray}
\fmerit_{2.2} \approx 2.2\times 1.9\times 10^{24}= 4.2\times 10^{24}\ \rm GeV\times pot, \\
\fmerit_{4.4} \approx 4.4\times 8.2\times 10^{23}= 3.6\times 10^{24}\ \rm  GeV\times pot, \\
\fmerit_{20} \approx 20\times 1.0\times 10^{23}= 2.0\times 10^{24}\ \rm  GeV\times pot, \\
\fmerit_{50} \approx 50\times 5.2\times 10^{22}= 2.6\times 10^{24}\ \rm  GeV\times pot, \\
\fmerit_{400} \approx 400\times 1.3\times 10^{22}= 5.2\times 10^{24}\ \rm  GeV\times pot.
\end{eqnarray}
This directly confirms the approximate scaling with $\fmerit$, apart at the lowest and the highest proton energies.
Strictly speaking, a proton energy of 20~GeV appears to be the most
economical choice in terms of protons, with a proton economy of about a factor 2 compared
to the 2.2 and 400~GeV cases.

Summarizing, the required protons on target for 5 oscillated
events at $\Delta m^2=3\times 10^{-3}\rm\ eV^2$ and $\sin^22\theta_{13}=10^{-3}$
in the 2.35~kton mass located {\it at the optimal baseline} assuming
a 20\% ideal focusing are $\approx 2\times 10^{24}$ for 2.2~GeV protons, 
$\approx 10^{24}$ for 4.4~GeV, $\approx 10^{23}$ for 20~GeV,
$\approx 5\times 10^{22}$ for 50~GeV and $\approx 10^{22}$ for 400~GeV.


\section{Results in real focusing}
\label{sec:realfoc}
The standard focusing system for all neutrino beams up to now is based on
 magnetic horns. However, new solutions should  be envisaged for low-energy, 
intense beams. The angular acceptance of the system has to be large 
(the average transverse momentum of reaction products 
is of the order of 200-300 MeV, comparable with the total momentum for
the shortest baselines), and the amount of material within the secondary beam 
cone must be as little as possible, both to preserve the flux and to avoid
 heating/damage at high intensities. 

In principle, the focusing system should cancel the transverse
momentum of all secondary particles relative to the direction of flight
toward the detector and this {\it independently of their momentum} $p$. 
Indeed, this is the definition of the ideal focusing.

In practice, a horn system can be designed to focalize a given signed
momentum, i.e. like $p$.  A FODO system focalizes like $|p|$. A series
of coils should focalize like $1$, i.e. independent of $p$. Hence, this last solution
seems to be attractive in our situation. 

We have found with the help of full
simulations and tracking of the particles in the focusing system
that (1) for the lowest energy configuration, a series of coils does indeed
provide a quite optimal focusing and (2) for the higher energy configurations,
where the coil focusing becomes impracticable, the traditional horn focusing
can be efficiently used.

\subsection{Details on coil focusing}
One possible solution is to exploit the focusing capabilities of magnetic
field gradients, such as the fringe field at the ends of a solenoid, or the
decreasing field far from a current loop. The fact that a particle
traveling almost parallel to the axis of a solenoid suffers a change of
its transverse momentum when traversing the fringe region is not intuitive
but can be understood by realizing that any change of the longitudinal
component of an axially symmetric field is associated through Maxwell's
divergence equation to a radial component of the field. 
 
The particle motion in axially symmetric fields can be described 
on the basis of the Busch's theorem, which
in turn follows from the conservation of canonical angular momentum.
The derivation and applications of this
theorem can be found in textbooks (see for instance \cite{Reiser,Jackson}), 
here we give a summary to 
better illustrate our focusing system. Suppose we have a particle having
electric charge $q$ and  total energy $E$, moving in 
a static, non-uniform axially symmetric field, at a distance 
 $r$ from the field axis ($z-$axis), with an angular velocity $\dot \phi$.
Busch's theorem states that 
\begin{equation}
r p_\phi + \frac{q}{2\pi}\Psi = const
\end{equation}
where  $p_\phi=\gamma m \dot \phi $ is the azimuthal component of the
particle momentum, and 
$\Psi$ is the magnetic flux linked by a circle of radius $r$ centered on
the field axis.      
 
If the field can be considered constant within the circle of radius $r$, 
we have 
\begin{equation}
\Psi = \pi r^2 B_z, \ \mathrm{thus} \ \ 
r p_\phi + \frac{q}{2} r^2 B_z = const 
\end{equation}

If particles are emitted with zero angular velocity by a source 
located on the axis of an axially symmetric magnetic field, both the
angular momentum and $\Psi$ are null, thus $const=0$.

When exiting to a
region of zero field, according to Busch's theorem the particles will have
{\it zero} angular momentum. This is very appealing remembering that in an uniform
(solenoidal) field, all the transverse momentum is azimuthal. However, this
is not the case when the field varies, and part of the momentum can also
be directed along the radius. The radial motion can also be
derived exploiting Busch's theorem, but the practical use is not 
straightforward. 
  
However, if the variation of the magnetic field is small on the scale of the
revolution time, the theorem of adiabatic invariance states that the
magnetic flux encircled by the particle trajectory remains a constant 
of motion. The trajectory radius can be assumed constant over a revolution
and given by the usual expression  $$R=\frac{p_T}{qB_z},$$ and from
adiabatic invariance follows $$B_z R^2\pi=  \frac{\pi p_T^2}{q^2B_z}=const.$$
Under this condition, all the rotational motion is converted
into the longitudinal one, and the particle transverse momentum varies as
\begin{equation}
 \frac{p^2_T}{B_z} =\frac{p^2_{T0}}{B_{z0}}
\end{equation}
 where $B_{z0}, \ p^2_{T0}$ are the initial 
magnetic field and transverse momentum. 

Since the period of the motion is 
$$\tau=\frac{2\pi}{\omega_B} =\frac{2\pi E}{q B_z},$$ the condition of
adiabatic motion means that $\Delta B_z$ must be small over distances
$$\Delta z = \frac{2\pi E v_\parallel}{q B_z} = \frac{2\pi  p_\parallel}{q B_z}.$$
 For a pion having a longitudinal momentum of 1 GeV/c, in  B=1 T, $\Delta
z$ is about 20 m. Is is thus difficult to focus efficiently high energy
particles through fringe fields, while this method can be very effective
for low energy ones. It should be stressed that this method applies equally
well to positive and negative particles, even though the resulting
advantage is not great since the positive component is dominant for low
energy beams.

\begin{table}[htb]

  \begin{center}
    \begin{tabular}{|c|c|c|c|c|c|c|c|c|}
\hline
\hline
Baseline $L$&\multicolumn{2}{c|}{2.2 GeV}
&\multicolumn{2}{c|}{4.4 GeV}
&\multicolumn{2}{c|}{20 GeV}
&\multicolumn{2}{c|}{400 GeV}\\
km
&\multicolumn{2}{c|}{coil focus}
&\multicolumn{2}{c|}{coil focus}
&\multicolumn{2}{c|}{horn focus}
&\multicolumn{2}{c|}{horn focus}\\
& $N^0_{\mu,CC}$ & $\inteni$
& $N^0_{\mu,CC}$ & $\inteni$
& $N^0_{\mu,CC}$ & $\inteni$
& $N^0_{\mu,CC}$ & $\inteni$ \\
\hline
 100  &  0.022  & $  9.1\cdot 10^{23}$
&  0.036  & $  5.6\cdot 10^{23}$
  &   0.10  & $  2.1\cdot 10^{23}$
  &    0.2  & $  1.0\cdot 10^{23}$
\\
 150  &  0.025  & ${\bf  8.0\cdot 10^{23}}$
&  0.053  & $  {\bf 3.8\cdot 10^{23}}$
  &   0.21  & $  9.5\cdot 10^{22}$
  &    0.6  & $  3.1\cdot 10^{22}$
\\
 200&  0.019  & $  1.0\cdot 10^{24}$  
&  0.053  & $  {\bf 3.8\cdot 10^{23}}$
  &   0.31  & $  6.5\cdot 10^{22}$
  &    1.4  & $  1.5\cdot 10^{22}$
\\
 300 &  0.009  & $  2.2\cdot 10^{24}$
&  0.027  & $  7.5\cdot 10^{23}$
  &   0.38  & $  5.3\cdot 10^{22}$
  &    2.0  & $  9.8\cdot 10^{21}$
\\
 400 
 &  0.005  & $  4.0\cdot 10^{24}$
 &  0.015  & $  1.3\cdot 10^{24}$
  &   0.39  & $  {\bf 5.2\cdot 10^{22}}$
  &    2.5  & $  8.0\cdot 10^{21}$
\\
 500  &  0.003  & $  6.4\cdot 10^{24}$
&  0.010  & $  2.1\cdot 10^{24}$
  &   0.38  & $  {\bf 5.2\cdot 10^{22}}$
  &    3.0  & $  6.8\cdot 10^{21}$
\\
 600  &  0.002  & $  9.8\cdot 10^{24}$
&  0.006  & $  3.1\cdot 10^{24}$
  &   0.38  & $  {\bf 5.2\cdot 10^{22}}$
  &    3.7  & $  5.4\cdot 10^{21}$
\\
 730  &  0.001  & $  1.6\cdot 10^{25}$
&  0.004  & $  4.9\cdot 10^{24}$
  &   0.35  & $  5.8\cdot 10^{22}$
  &    3.8  & $  {\bf 5.2\cdot 10^{21}}$
\\
\hline
   \end{tabular}
  \end{center}
\caption{Same as Table \protect\ref{tab:rateosci} with
the focusing system included in the simulations. The configuration corresponding to the minimum
$\inteni$ is shown in bold.
It is computed in order to obtain a number of oscillated events
$N_e=5$ (see text) for the various baselines $L$ and
proton energies. 
}
\label{tab:ratefoc}
\end{table}

\subsection{Results on horn and coil focusing}
We report in Table~\ref{tab:ratefoc} our results with full detailed 
simulations of the focusing
systems for 2.2~GeV 4.4~GeV, 20~GeV and 400 GeV proton energies.

\begin{figure}[p]
\centering
\epsfig{file=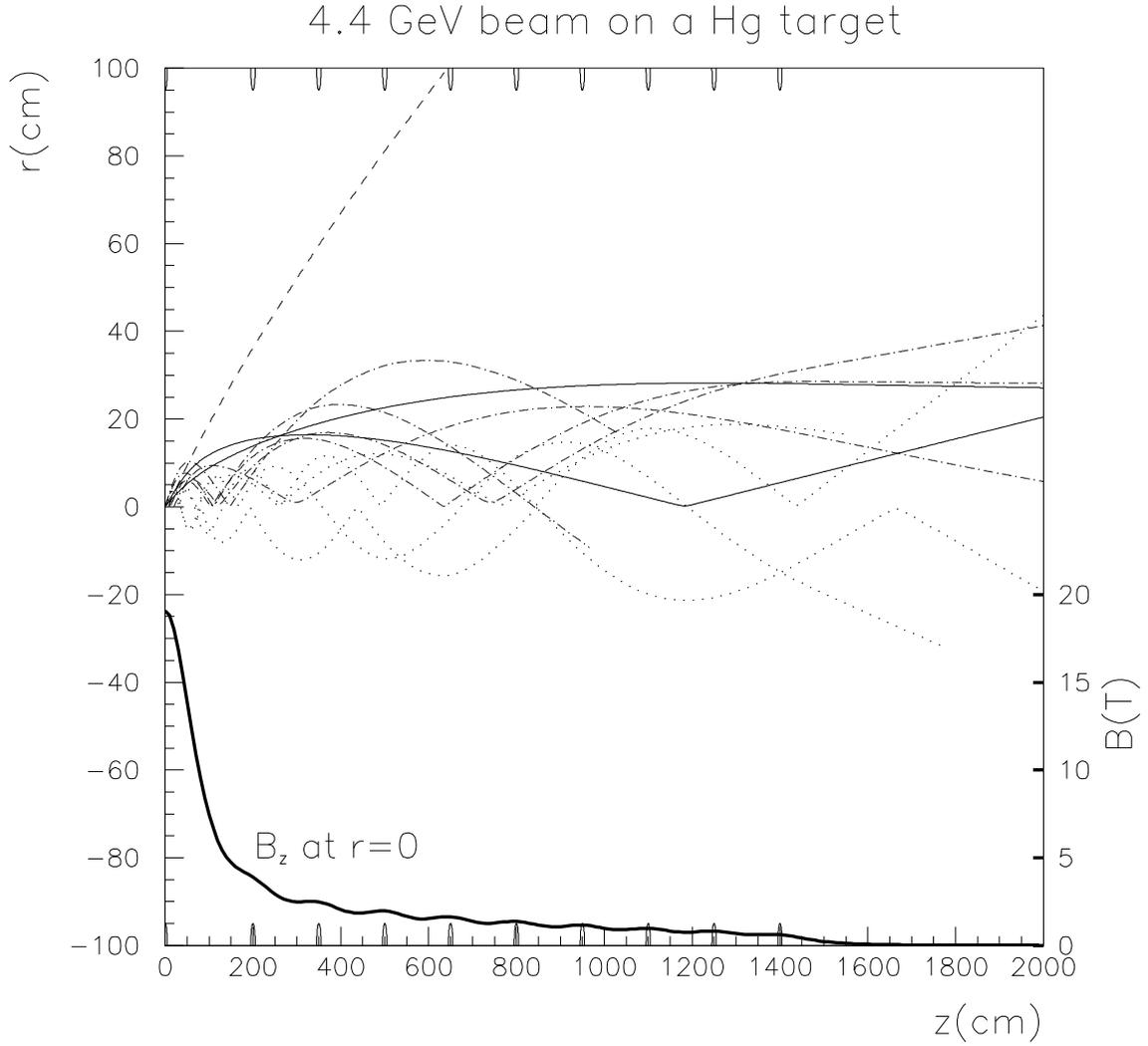,%
width=15cm%
,bbllx=0.pt,bblly=30pt,bburx=510pt,bbury=520pt%
}
\caption{ Particle trajectories in the current loops focusing, from a 4.4
GeV proton beam on a Hg target. Dotted tracks: $p_\pi < 0.5 $ GeV,
dot-dashed tracks $0.5 < p_\pi < 1 $ GeV,
continuous tracks $1 < p_\pi < 2 $ GeV,
dashed tracks $p_\pi > 2 $ GeV. The thick curve is the on-axis 
magnetic field value (scale on the right). Loop positions are marked on top
and bottom.
}
\label{fig:coiltraj}
\end{figure}

\begin{figure}[p]
\centering
\epsfig{file=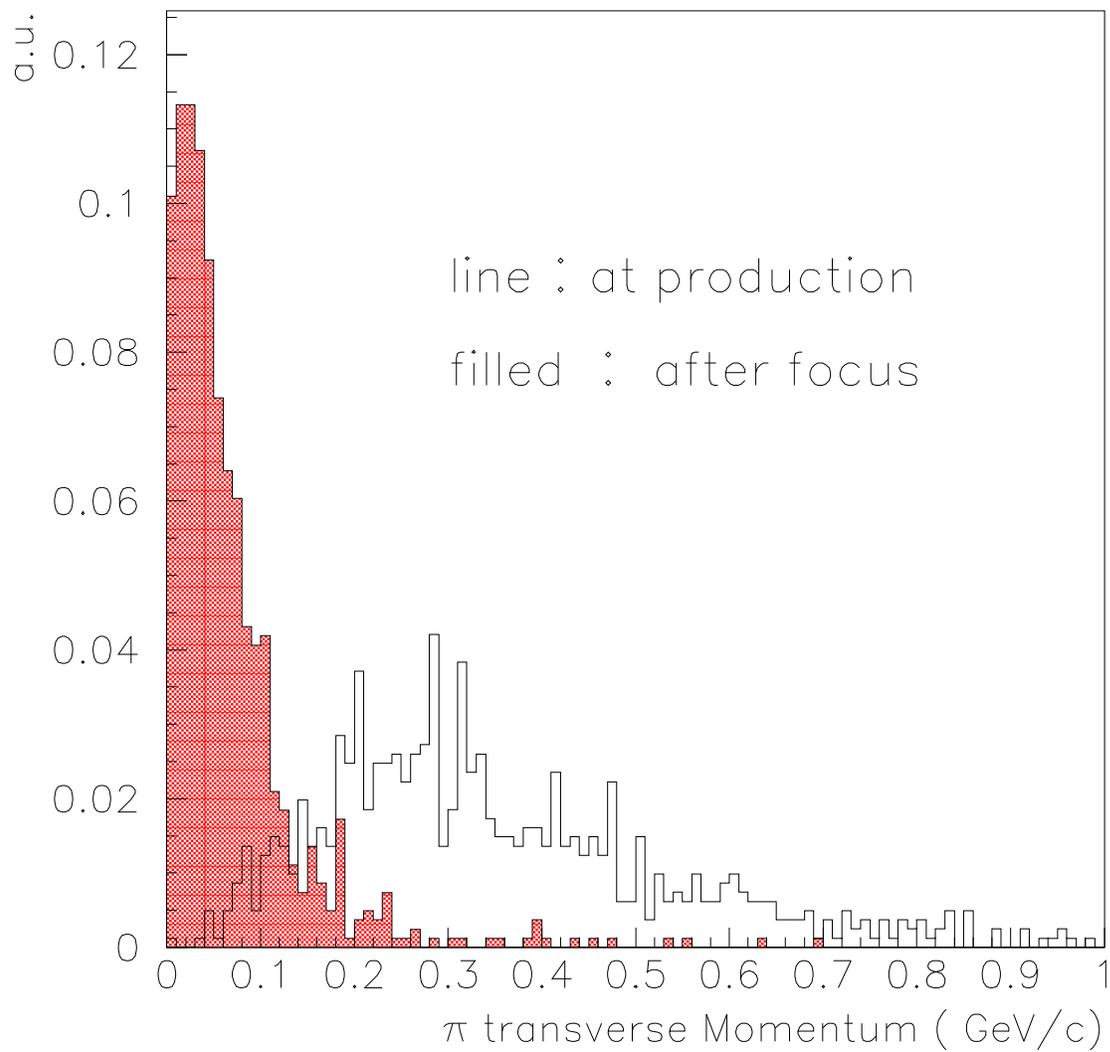,width=15cm%
,bbllx=30pt,bblly=180pt,bburx=500pt,bbury=651pt}
\caption{Effect of a multiple current loop system on the transverse
momentum distribution, in the case of a 4.4 GeV beam on a mercury target}

\label{fig:foc4}
\end{figure}

The coil method has been applied here to the 2.2 and  4.4~GeV proton
beams, where the produced pions have  small energies (only positive mesons
have been considered in Table~\ref{tab:ratefoc}).
The target has been assumed to be
a short (30~cm) mercury target, like in the SPL\cite{SPL} proposal. 
The non-uniform magnetic field has been obtained with ten circular loops,
having a radius of 1 m,
positioned  from 0 to 14 meters from the target, carrying decreasing
currents to give a field from 20~T to zero. Examples of particle orbits and 
the central magnetic field
intensity are shown in Figure~\ref{fig:coiltraj}. The effect on the transverse
momentum distribution can be appreciated from Figure~\ref{fig:foc4}, where 
positive pions emitted from the target within 1 radian have been considered.
The decay tunnel is 150~m long.

\begin{figure}[p]
\centering
\epsfig{file=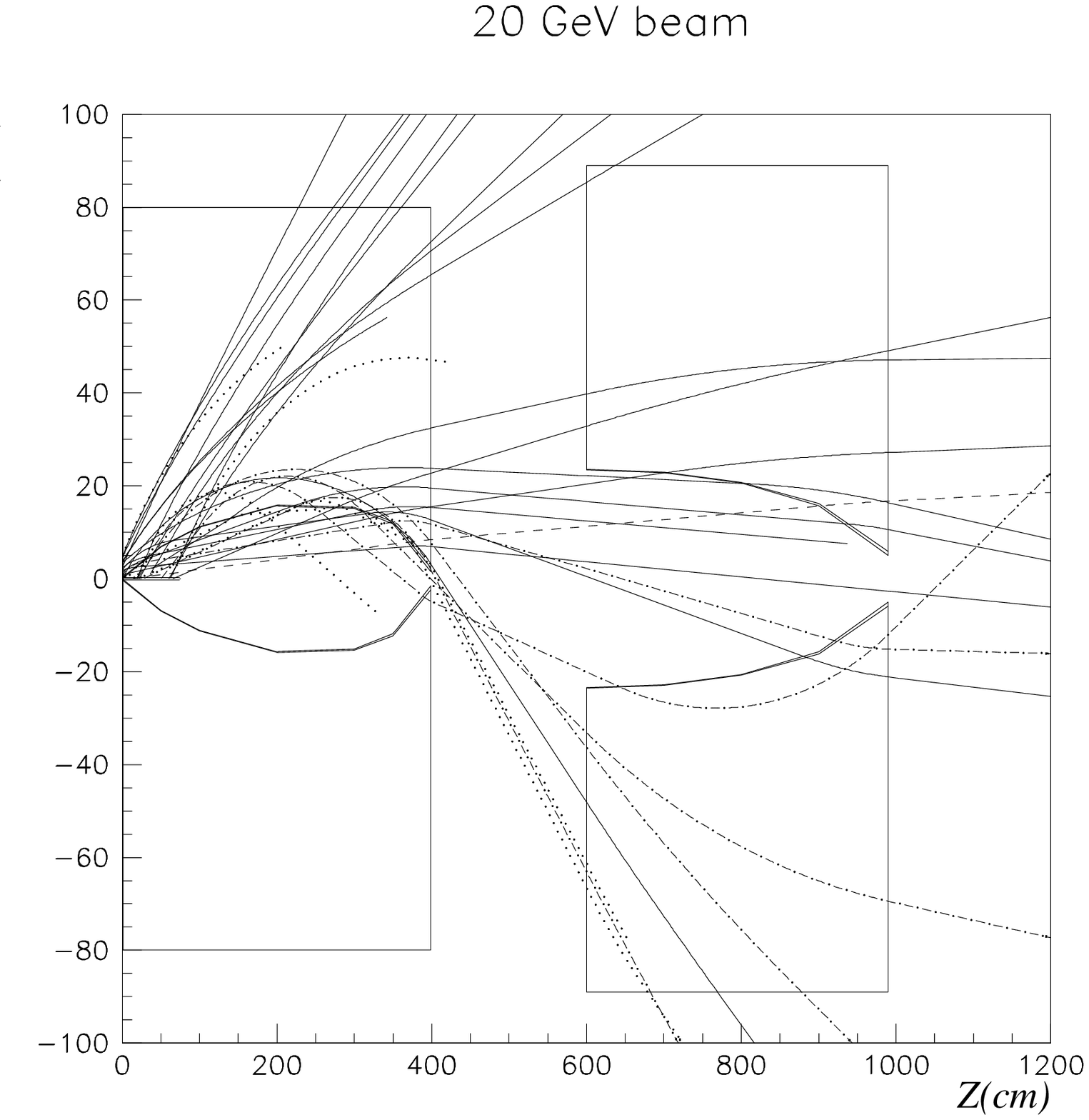,%
width=15cm%
,bbllx=0.pt,bblly=30pt,bburx=480pt,bbury=520pt%
}
\caption{ Particle trajectories in the horn+reflector focusing, from a 20
GeV proton beam on a C target. Dotted tracks: $p_\pi < 0.5 $ GeV,
dot-dashed tracks $0.5 < p_\pi < 1 $ GeV,
continuous tracks $1 < p_\pi < 6 $ GeV,
dashed tracks $p_\pi > 6 $ GeV. 
}
\label{fig:horntraj}
\end{figure}

\begin{figure}[p]
\centering
\epsfig{file=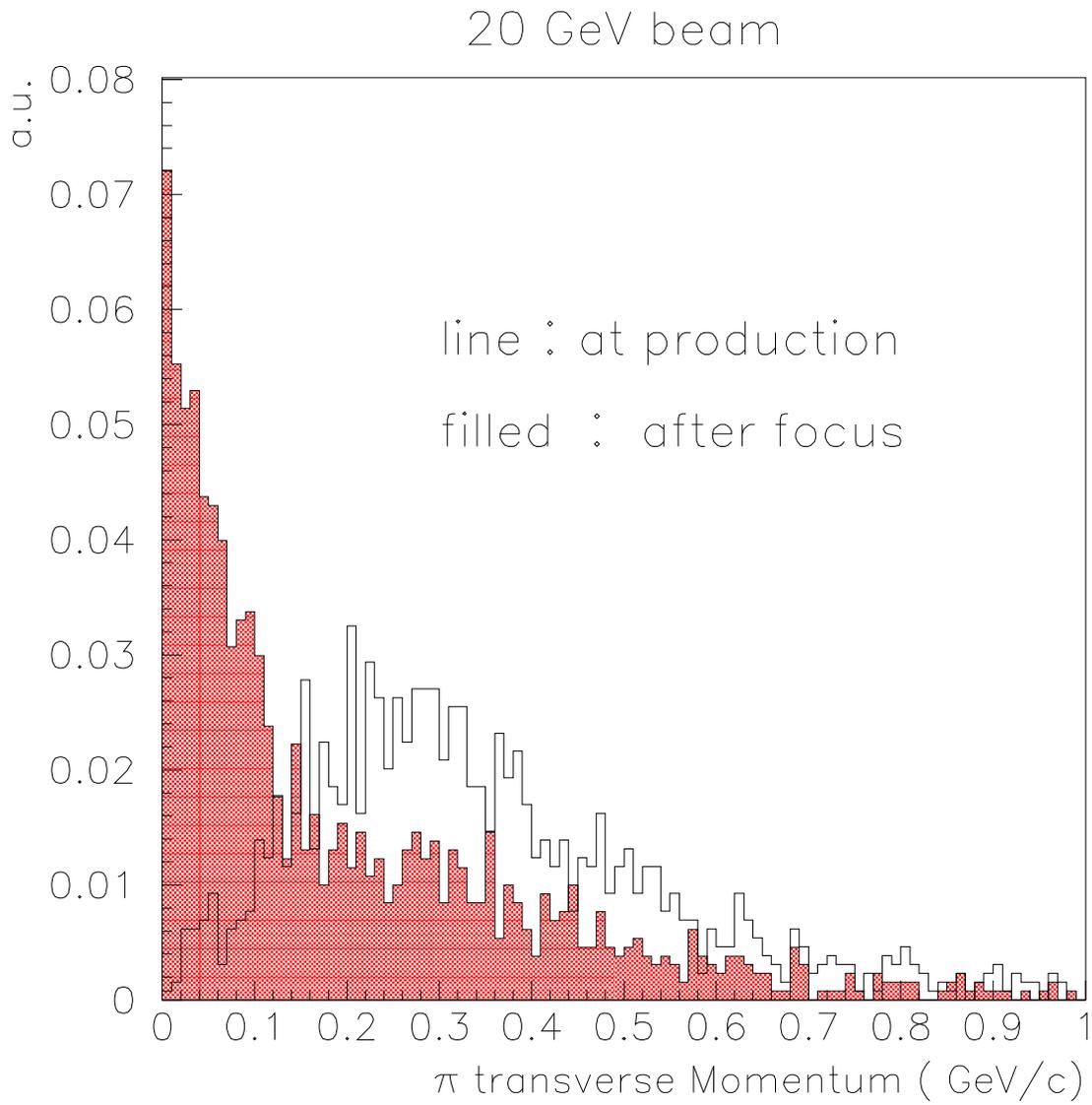,width=15cm%
,bbllx=30pt,bblly=180pt,bburx=500pt,bbury=651pt}
\caption{Effect of a double horn focusing on the transverse
momentum distribution, in the case of a 20 GeV beam on a carbon target}

\label{fig:fhorn}
\end{figure}

For higher energy neutrino beams of 20~GeV and 400 GeV, the traditional 
two-horns system has been used. The calculations presented here refer to a first horn to
focus 2 GeV/c particles, followed by a reflector to focus 3 GeV/c. The
horn is  placed
around a graphite target, has a length of 4 meters and a current of 300~kA. The
reflector starts at 6~m from the target, is 4 m long with a current of
150~kA. 
Examples of particle trajectories can be seen in 
Figure~\ref{fig:horntraj}, and 
the effect on the transverse momentum in the
case of a 20 GeV primary beam is shown in Figure~\ref{fig:fhorn}. 
The decay tunnel is 350~m long.

In the total energy range (0-2.5 GeV), the resulting
focusing efficiency varies in between 0.5 and
0.2. This motivated our choice in Section~\ref{sec:ideafoc}, when comparing all 
possible beam/distance options, 
where a common energy-independent
20\% focusing efficiency  had been assumed. In reality, the
focusing profile is not energy independent, and the real situation can be
better than this, as can be derived
from the comparison of Tables ~\ref{tab:rateosci} and \ref{tab:ratefoc}. 
This effect is also visible in Figure~\ref{fig:rateosci}.

Summarizing, the required protons on target for 5 oscillated
events at $\Delta m^2=3\times 10^{-3}\rm\ eV^2$ and $\sin^22\theta_{13}=10^{-3}$
in the 2.35~kton mass located {\it at the optimal baseline} assuming
optimized realistic focusing are $\approx 1\times 10^{24}$ for 2.2~GeV protons, 
$\approx 4\times 10^{23}$ for 4.4~GeV, $\approx 5\times 10^{22}$ for 20~GeV
and $\approx 5\times 10^{21}$ for 400~GeV.

The cautious reader can wonder why we have apparently been ``conservative'' in
assuming a 20\% efficiency for the ideal focusing. The point here is that in the ideal case
we assume a constant efficiency over the whole meson energy range, while in the real
focusing case one effectively reaches a situation where a part of the meson energy
range is focalized with an efficiency better than 20\% while other parts have lower
efficiencies. We have of course optimized focusing for the energy relevant to the oscillation.
It would however be incorrect to assume that this higher efficiency is constant over
the full energy range.


\begin{table}[tb]
\begin{center}
\begin{tabular}{|l||c|c|c|c|c|}
\hline
Laboratory&Lat.&Long.&Baseline &Angle&Inclination\\
&&&to LNGS (km)&&\\
\hline
Casaccia (ENEA)&42$^\circ$&12.2$^\circ$& {\bf 120} & 63$^\circ$& 0.54$^\circ$\\
Napoli (ENEA) &40.8$^\circ$&14.3$^\circ$& 200&145$^\circ$& 0.9$^\circ$\\
Aquilone (ENEA) &41.7$^\circ$&15.9$^\circ$& 217&166$^\circ$& 1.$^\circ$\\
Brasimone (ENEA) &44.2$^\circ$&11.1$^\circ$& {\bf 270} &8$^\circ$& 1.2$^\circ$\\
Legnaro (INFN) &45.4$^\circ$&12.0$^\circ$& 344&34$^\circ$& 1.5$^\circ$\\
Trisaia (ENEA) &40.3$^\circ$&16.8$^\circ$& 368&173$^\circ$& 1.65$^\circ$\\
Pavia (INFN,Univ.) &45.2$^\circ$&9.2$^\circ$&{\bf 465}&5.7$^\circ$& 2.1$^\circ$\\
Ispra (ENEA,CCR) &45.8$^\circ$&8.6$^\circ$& 536&8.5$^\circ$& 2.4$^\circ$\\
Catania (INFN) &37.5$^\circ$&15.1$^\circ$& 570&140$^\circ$& 1.5$^\circ$\\
\hline
CERN &46.1$^\circ$&6.0$^\circ$& 732&0.0$^\circ$& 3.22$^\circ$\\
\hline
\end{tabular}
\end{center}
\caption{List of existing laboratories with interesting
baselines to the LNGS Gran Sasso
laboratory. Angle is the space angle relative to the orientation
of the LNGS Halls. Inclination is the incoming neutrino angle relative
to the horizontal plane at LNGS.}
\label{tab:mediumbaseline}
\end{table}

\section{A superbeam to Gran Sasso (BNGS\protect\footnote{BNGS stands for ``Better Neutrinos to GS''.}) ?}
\label{sec:bngs}

\subsection{Finding the location for the source}

As a working hypothesis, we take for granted the existence of
LNGS as an underground laboratory that can host large
neutrino detectors. In particular, we have assumed for this
study the existence of
ICARUS with 2.35~kton fiducial mass.

If we take the location of the detector as fixed, the baseline $L$
is determined by the location of the neutrino source. We have investigated
various potential locations within Italy where large
ENEA\footnote{ENEA stands for ``Ente per le Nuove Tecnologie, l'Energia
e l'Ambiente''.} or INFN infrastructures are already existing, as shown 
in Table~\ref{tab:mediumbaseline}.
In these locations, the required conditions to host a high
intensity proton driver could be met and the machine would find
other applications in addition to neutrino physics. 

In Table~\ref{tab:mediumbaseline}, the column called ``angle'' 
describes the space angle relative to the orientation
of the LNGS Halls. Inclination is the incoming neutrino angle relative
to the horizontal plane at LNGS.
We note that due to the fortunate orientation of
the LNGS Halls according to the geographical axis of the Italian peninsula,
it is possible to find various laboratories within the orientation of
the LNGS Halls. 

Indeed, ENEA Aquilone, ENEA Brasimone, ENEA Trisaia,
INFN Pavia and ENEA Ispra appear
with an angle less than $15^o$ with respect to the LNGS Hall
direction. This is an advantage for the acceptance of higher energy events
given the natural longitudinal orientation of the detectors. On the other
hand, in the case of ENEA Casaccia situated near Rome,
the angle is $63^o$. However, for the shortest baseline, we expect the 
relevant neutrinos to have an energy
similar to that of most atmospheric neutrinos, so that we can argue that the acceptance
will not be a problem given the isotropical nature of a detector like ICARUS.


\begin{table}[tb]
\begin{center}
\begin{tabular}{|c|c|c|c|c|c|c|c|c|c|}
\hline
&&&&&\multicolumn{2}{c|}{ $10^{23}$ p.o.t. 
}&\multicolumn{2}{c|}{$<E_\nu>$, CC}&\\
$E_p$&Focus&Decay tunnel & $\nu_\mu$ flux & $\nu_e$ flux&$\nu_\mu$ CC & $\nu_e$ CC 
&$\nu_\mu$ &$\nu_ e$ & $\nu_e / \nu_\mu$ \\  
GeV&&length (m)&\multicolumn{2}{c|}{$\nu$/cm$^2$}&
\multicolumn{2}{c|}{events/kton}&
\multicolumn{2}{c|}{GeV }&CC\\
& & & & & & & & &\\
\hline
 2.2 & p.f.& 150 & 1.6 $\times10^{-11}$&8.0 $\times10^{-13}$&285&4.9&0.47& 0.53&1.7\%\\
2.2 & coil& 150 & 9.5 $\times10^{-12}$&1.2 $\times10^{-13}$&109& 2.2 &0.41&0.56&2.1\%\\
\hline
4.4 & p.f.& 150 & 5.7 $\times10^{-11}$&1.6 $\times10^{-12}$&1900&28&0.77&0.96&1.5\%\\
4.4 & coil& 150 & 2.0 $\times10^{-11}$&9.5 $\times10^{-13}$&340&5.6&0.56&0.6&1.6\% \\
\hline
20& p.f.& 350 & 1.1$\times10^{-9}$&2.3 $\times10^{-11}$&6.6$\times10^{4}$&800&1.6&1.4&1.3\%\\
20& horn& 350 & 4.4$\times10^{-10}$&7.3$\times10^{-12}$&2.6$\times10^{4}$&310&1.6&1.3&1.2\%\\
\hline
\hline
\end{tabular}
\end{center}
\caption{Neutrino beam parameters for a 270 km baseline 
experiment. All quantities are calculated for $0< E_\nu <2.5\rm\ GeV$}
\label{tab:Brasimone}
\end{table}

\subsection{The medium baselines ($120<L<470\rm\ km$)} 

We study the cases ENEA Casaccia ($L=120\rm\ km$), 
ENEA Brasimone ($L=270\rm\ km$) and INFN Pavia ($L=465\rm\ km$). 
At these distances, the neutrino energy range relevant for $\nu_\mu
\rightarrow \nu_e$ search is 0.1-1.0 GeV.  
We assume 2.2, 4.4 and 20~GeV proton energies.
Expected neutrino fluxes and rates, obtained assuming ideal (p.f.) and
real (coil or horn) focusing systems, are reported in Table~\ref{tab:Brasimone}. The
figures are normalized to
the baseline $L=270\rm\ km$ and for $0\leq E_\nu\leq 2.5\rm\ GeV$. 
Other baselines can be rescaled accordingly.
With real focusing, we find about 100(10) $\numu$($\bar\numu$) CC events/kton per $10^{23}$~pots
for 2.2~GeV protons, about 340(40) $\numu$($\bar\numu$) CC events/kton for 4.4~GeV protons
and about 26000(250) $\numu$($\bar\numu$) CC events/kton for 20~GeV protons.
Intrinsic $\nue$($\bar\nue$) beam contaminations are in the range from 1\% to 2\%
with respect to $\numu$($\bar\numu$).

The 2.2 and 4.4~GeV proton energies are the closest to the optimal energies
and only these two cases are considered in the following. We will consider 20~GeV
in section~\ref{sec:cerngs} in the context of the CERN-LNGS baseline.

\begin{figure}[p]
\centering
\epsfig{file=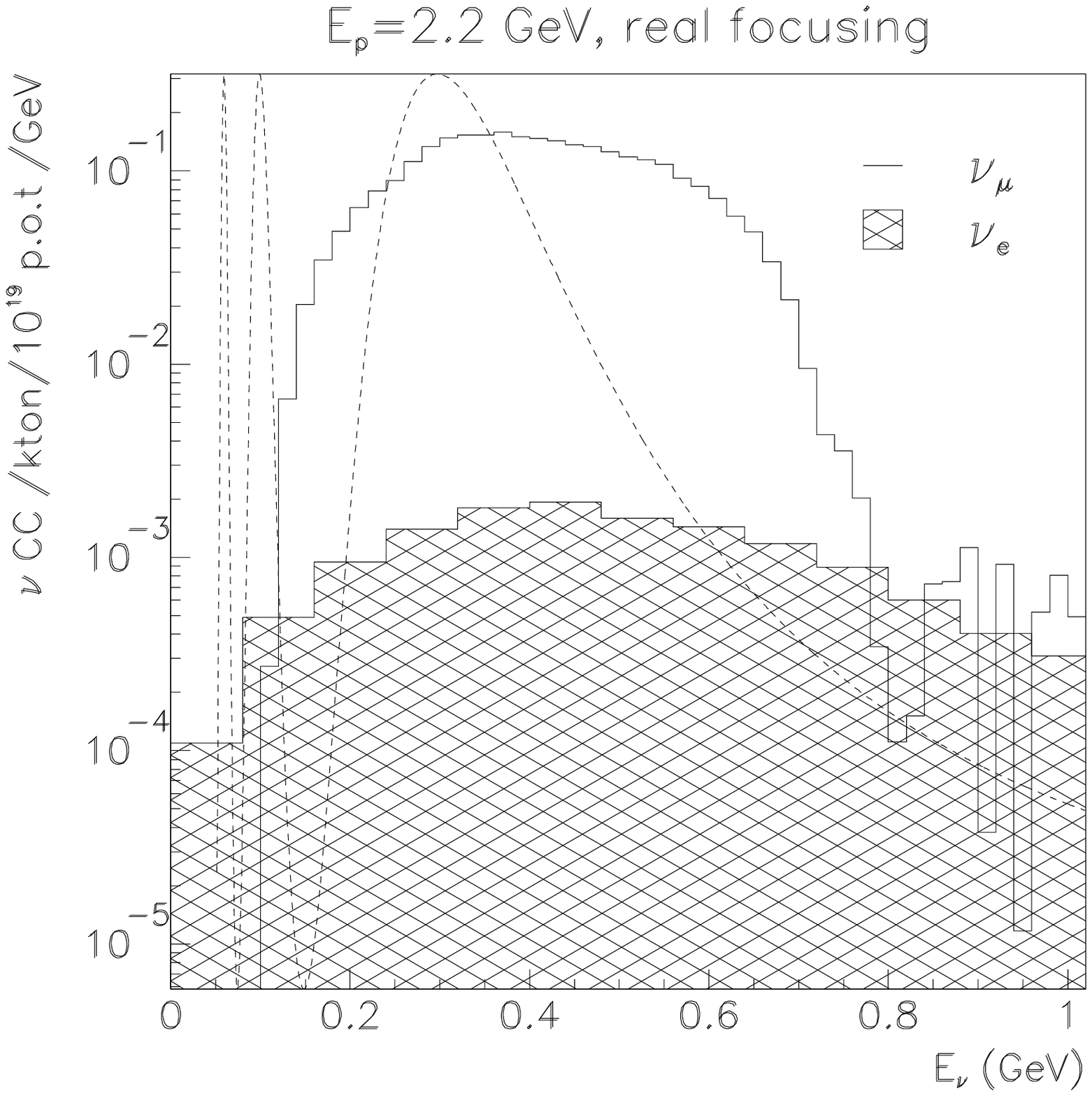,width=15cm}
\caption{Charged current event rate at 120 km as a function of the neutrino energy
computed with (real) coil focusing for 2.2~GeV proton energy. The dotted line corresponds
to the oscillation probability (arb. norm.) for a $\Delta m^2=3\times 10^{-3}\rm\ eV^2$.}  
\label{fig:plotenespl_e}
\end{figure}

\begin{figure}[p]
\centering
\epsfig{file=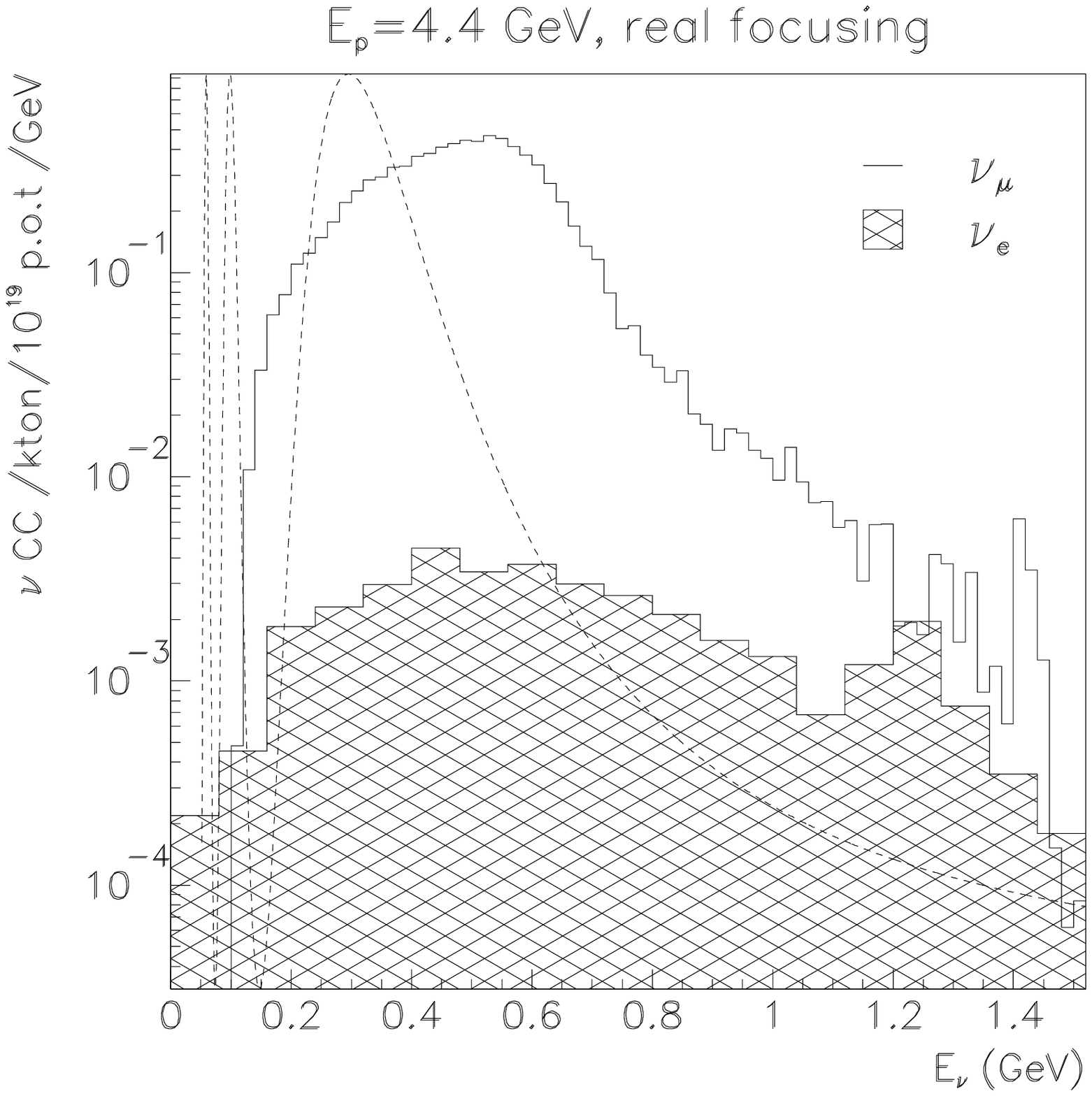,width=15cm}
\caption{ 
Charged current event rate at 120 km as a function of the neutrino energy
computed with (real) coil focusing for 4.4~GeV proton energy. The dotted line corresponds
to the oscillation probability (arb. norm.) for a $\Delta m^2=3\times 10^{-3}\rm\ eV^2$.}
\label{fig:plotenesplf_e}
\end{figure}

To appreciate the matching between the neutrino beams and the oscillation
probability, we show in Figures~\ref{fig:plotenespl_e} and \ref{fig:plotenesplf_e}
the charged current event rate at 120 km as a function of the neutrino energy. 
The dotted lines correspond to the oscillation probability (arb. norm.) for a 
$\Delta m^2=3\times 10^{-3}\rm\ eV^2$.

Clearly, these neutrino beams offer optimal condition to study $\numu$ {\it disappearance}.
Indeed, the maximum of the oscillation is very well covered by the neutrino  beam and hence
it is quite obvious that a very precise determination of the main oscillation parameters 
will be accomplished. We here do not consider this any further.

We, however, concentrate instead on the $\nue$ {\it appearance} measurement. Since 
we are in the presence of intrinsic $\nue$ background from the beam at the level of
1\%-2\% of the $\numu$ component, we can improve our sensitivity by studying the
energy spectrum of the $\nue$ charged current events. This method is more sensitive
than simple event counting.

In order to estimate the sensitivity, we adopt our standard fitting procedure of
the various reconstructed event classes (See Ref.~\cite{Bueno:2000fg}).
We assume that the neutrino and antineutrino interactions cannot be distinguished
on an event-by-event basis, and hence add the $\nue$ and $\bar\nue$ contributions from the beam. Similarly,
the oscillated spectrum is calculated by summing both $\numu\ra\nue$ and $\bar\numu
\ra\bar\nue$ oscillations, assuming the same oscillation probability for
neutrino and antineutrinos.

In the present study, we considered only the energy distribution of electron events
and computed the $\chi^2$ as a function of the $\sin^2 2\theta_{13}$ mixing angle,
scanning in $\Delta m^2$. The 90\% C.L. sensivity region is defined by the condition
$\chi^2>\chi^2_{min}+4.6$, defined by the condition that the actually observed events
in the experiment coincide with the expected background. In an actual experiment,
a simultaneous fit of the muon disappearance and electron appearance 
spectra will constrain the $\Delta m^2$, $\sin^2\theta_{23}$ parameters and in case of
negative result will limit the $\sin^2 2\theta_{13}$ within the allowed $\Delta m^2$
region.

\begin{figure}[p]
\centering
\epsfig{file=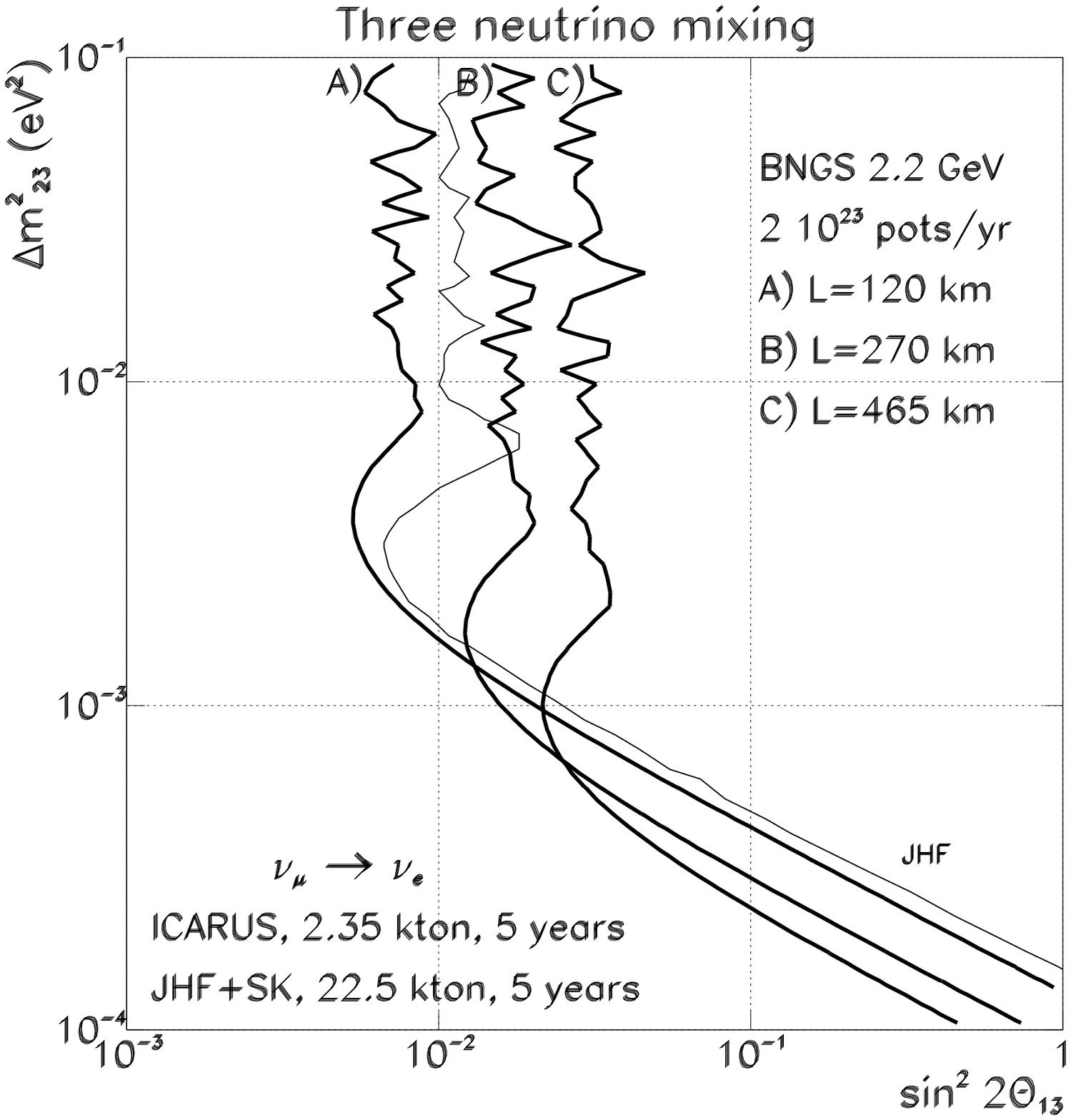,width=17cm}
\caption{Expected 90\% C.L. sensitivity 
to $\sin^22\theta_{13}$ mixing angle at 2.2 GeV proton energy
for the three possible baselines $L=120$, 270 and 465~km (thick lines), 
compared to expected results from JHF-SK\cite{JHF} (thin line).}
\label{fig:exclus2p2}
\end{figure}

\begin{figure}[p]
\centering
\epsfig{file=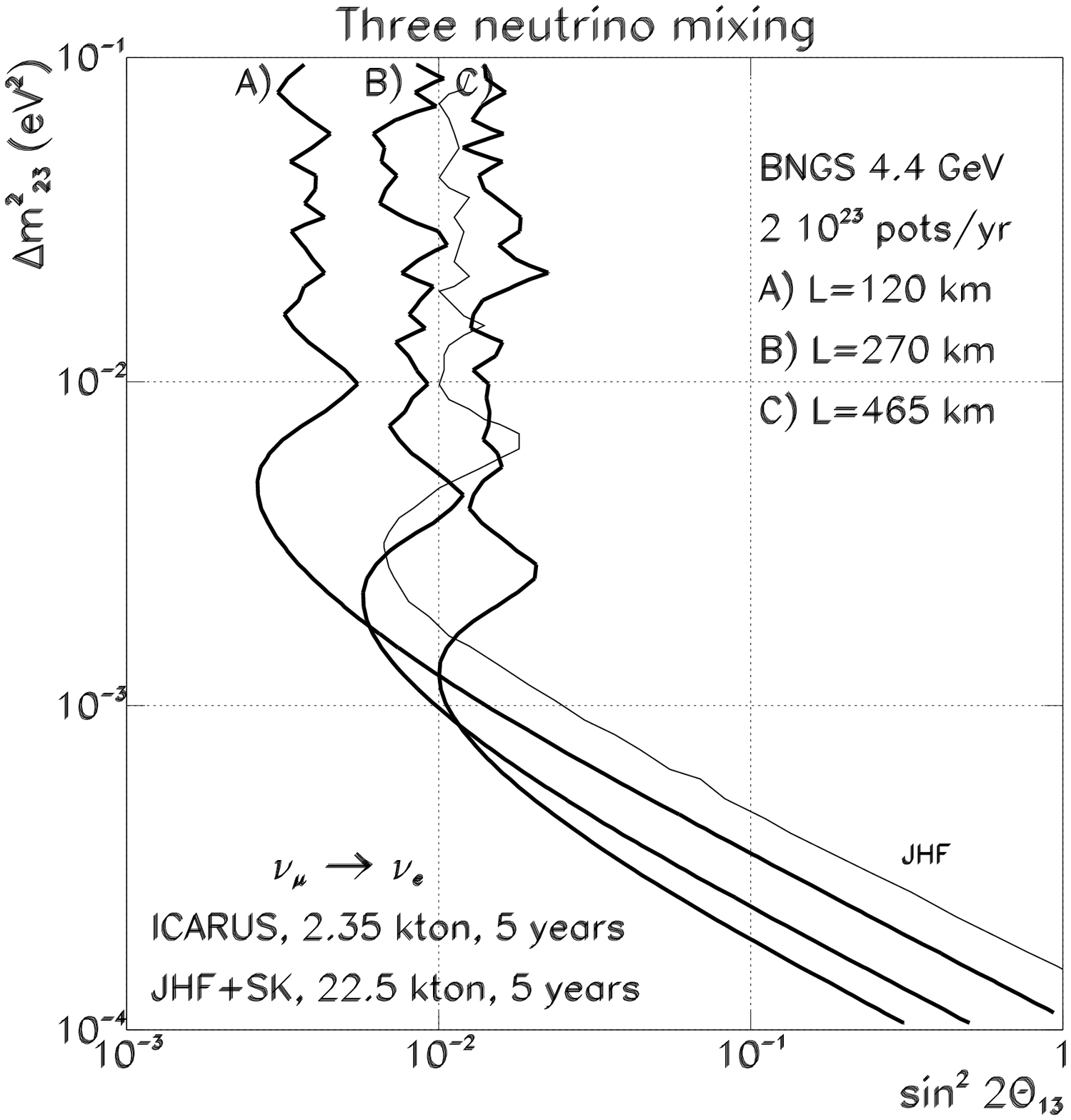,width=17cm}
\caption{Expected 90\% C.L. sensitivity 
to $\sin^22\theta_{13}$ mixing angle at 4.4 GeV proton energy
for the three possible baselines $L=120$, 270 and 465~km (thick lines), 
compared to expected results from JHF-SK\cite{JHF} (thin line).}
\label{fig:exclus4p4}
\end{figure}

The results of the $\Delta m^2$ scans 
are shown in Figure~\ref{fig:exclus2p2} for 2.2~GeV and 
Figure~\ref{fig:exclus4p4} for 4.4 GeV proton
energy for the three assumed baselines. 
The curves correspond to 5 years running with $2\times 10^{23}$ pots/year equivalent
to a continous proton current of 1~mA  and
a fiducial mass of 2.35~kton. The assumed protons on target is compatible with an
accelerator with performances similar to those of the planned CERN-SPL\cite{SPL}. 

The expected sensitivities represent a great improvement to
the CHOOZ limit\cite{Chooz} which gives
for $\Delta m^2=2.5\times 10^{-3}\rm\ eV^2$,
\begin{eqnarray}
(\sin^22\theta_{13})_{CHOOZ}< 0.14 \rm\ \ \ or\ \ \  \theta_{13}<11^o
\end{eqnarray}
Indeed, we find for $2.2\rm\ GeV$ protons:
\begin{eqnarray}
(\sin^22\theta_{13})_{BNGS,120km}& <& 0.006 \rm\ \ \ or\ \ \  \theta_{13}<2.2^o \\
(\sin^22\theta_{13})_{BNGS,270km}&< &0.015 \rm\ \ \ or\ \ \  \theta_{13}<3.4^o \\
(\sin^22\theta_{13})_{BNGS,465km}&< &0.03 \rm\ \ \ or\ \ \  \theta_{13}<5^o 
\end{eqnarray}
and for $4.4\rm\ GeV$ protons:
\begin{eqnarray}
(\sin^22\theta_{13})_{BNGS,120km}&<& 0.0035 \rm\ \ \ or\ \ \  \theta_{13}<1.7^o \\
(\sin^22\theta_{13})_{BNGS,270km}&<& 0.006 \rm\ \ \ or\ \ \  \theta_{13}<2.2^o \\
(\sin^22\theta_{13})_{BNGS,465km}&<& 0.02 \rm\ \ \ or\ \ \  \theta_{13}<4^o 
\end{eqnarray}
For comparison, the JHF proposal with the OAB beam gives similar results\cite{JHF}
\begin{eqnarray}
(\sin^22\theta_{13})_{JHF,OAB}< 0.006 \rm\ \ \ or\ \ \  \theta_{13}<2.2^o.
\end{eqnarray}


\subsection{The CERN-GS baseline} 
\label{sec:cerngs}

The baseline between CERN and GS
is 730 km. At this distance, the neutrino energy range relevant for $\nu_\mu
\rightarrow \nu_e$ search is 0.3-2.5 GeV.    

The present CNGS design\cite{CNGS} is optimized for 
$\nu_\tau$ appearance, thus for a 
relatively high-energy neutrino beam. The 400 GeV/c SPS  beam will
deliver 4.5 10$^{19}$ protons per year on  a
graphite target, made of spaced rods to reduce the re-interaction rate.
The two magnetic horns (horn and reflector) are tuned to focus 35 and 50
GeV/c mesons, with an acceptance of the order of 30~mrad. The decay tunnel 
length is 1 km.
 With the standard CNGS parameters, the low-energy neutrino flux is low, as can
be seen from the entries flagged by $^\dagger$ in Table~\ref{tab:CNGS}.

In Ref.~\cite{Rubbia:2002rb}, we have studied a L.E. optimization of the 400 GeV protons
of the CNGS in order to improve the sensitivity to $\theta_{13}$. This yielded
an improvement of a factor 5 in flux at low energy compared to the $\tau$
optimization.

Here, we study the 20~GeV proton energy (we call this
the PS++) and compare it to 400~GeV. 
Expected neutrino fluxes and rates obtained with real focusing systems
are reported in Table~\ref{tab:CNGS}. 

\begin{figure}[p]
\centering
\epsfig{file=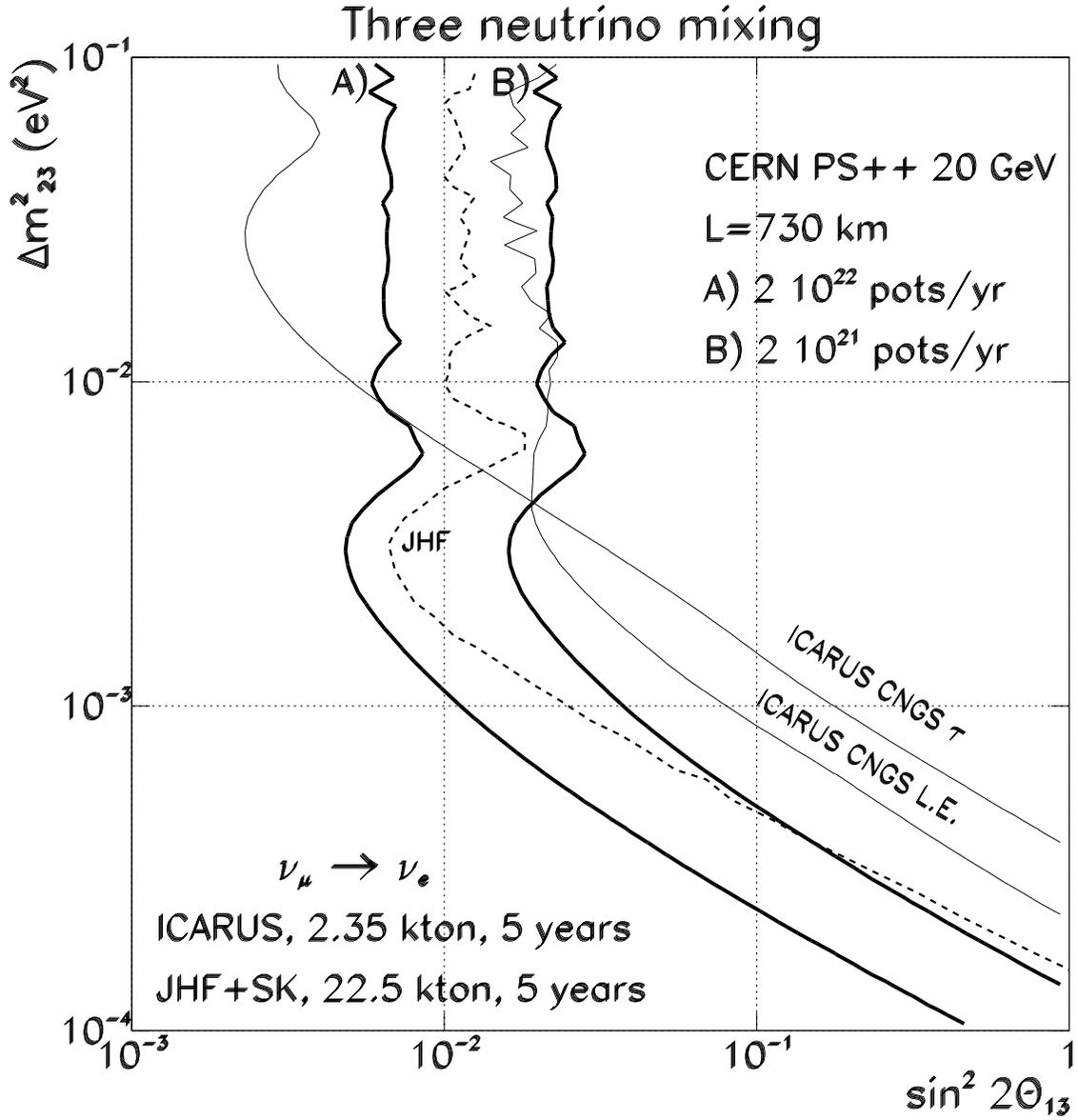,width=17cm}
\caption{Expected 90\% C.L. sensitivity 
to $\sin^22\theta_{13}$ mixing angle at 20 GeV proton energy
for the CERN-LNGS baseline (L=730~km) (thick lines) for two
yearly integrated proton intensities, 
compared to expected results from CNGS $\tau$ and L.E. optimized (See Ref.\cite{Rubbia:2002rb})
(thin lines)
and JHF-SK\cite{JHF} (dotted line).}
\label{fig:exclusp20}
\end{figure}

The results of the $\Delta m^2-\sin^22\theta_{13}$ sensitivity scans 
are shown in Figure~\ref{fig:exclusp20}. 
The curves correspond to 5 years running with $2\times 10^{21}$ or
$2\times 10^{22}$  pots/year and
a fiducial mass of 2.35~kton. 

We find for $\Delta m^2=2.5\times 10^{-3}\rm\ eV^2$:
\begin{eqnarray}
(\sin^22\theta_{13})_{PS++,2\times 10^{21} pot/year}< 0.016 \rm\ \ \ or\ \ \  \theta_{13}<3.6^o \\
(\sin^22\theta_{13})_{PS++,2\times 10^{22} pot/year}< 0.005 \rm\ \ \ or\ \ \  \theta_{13}<2^o
\end{eqnarray}

Clearly, a strongly intensity-upgraded CERN PS booster would provide very interesting possibilities
for the oscillation searches over the CERN-GS baseline.

\tabcolsep=.8mm
\begin{table}[htb]
\begin{tabular}{|c|c|c|c|c|c|c|c|c|c|}
\hline
&&&&&\multicolumn{2}{c|}{ $10^{19}$ p.o.t. 
}&\multicolumn{2}{c|}{$<E_\nu>$, CC}&\\
$E_p$&focus&Decay tunnel &$\nu_\mu$ flux & $\nu_e$ flux&$\nu_\mu$ CC & $\nu_e$ CC 
&$\nu_\mu$ &$\nu_ e$ & $\nu_e / \nu_\mu$ \\  
GeV&& length (m)&\multicolumn{2}{c|}{$\nu$/cm$^2$}&
\multicolumn{2}{c|}{events/kton}&
\multicolumn{2}{c|}{GeV }&CC\\
& & & & & & & & &\\
\hline
20 & p.f.&150
	&$ 9.8\times 10^{-15}$
	&$ 1.6\times 10^{-16}$
	&$ 0.56$
	&$ 4.1\times 10^{-3}$
&1.5&1.26&0.7\%\\
20 & p.f.&350
	&$ 1.5\times 10^{-14}$
	&$ 3.1\times 10^{-16}$
	&$ 0.9$
	&$ 0.011$
&1.6&1.4&1.3\%\\
20 &horn&350
	&$ 6.1\times 10^{-15}$
	&$ 1.0\times 10^{-16}$
	&$ 0.36$
	&$ 4.2\times 10^{-3}$
&1.6&1.3&1.2\%\\
\hline
400 & p.f  &350
	&$ 1.3\times 10^{-13}$
	&$ 2.6\times 10^{-15}$
	&$ 9.0$
	&$ 0.12$
&1.8&1.8&1.3\% \\
400 & horn &350
	&$ 1.0\times 10^{-15}$
	&$ 9.0\times 10^{-16}$
	&$ 4.5$
	&$ 4.2\times 10^{-2}$
&1.8&1.4&0.9\% \\
400 & p.f $^\dagger$&CNGS
	&$ 1.6\times 10^{-14}$
	&$ 3.2\times 10^{-16}$
	&$ 1.8$
	&$ 2.2\times 10^{-2}$
&2.1&1.7&1.2\% \\
400 & $\tau ^\dagger$&CNGS
	&$ 1\times 10^{-14}$
	&$ 9.4\times 10^{-17}$
	&$ 0.9$
	&$ 8.7\times 10^{-3}$
&1.8&1.8&0.9\%\\

\hline
\hline
\end{tabular}
\caption{Neutrino beam parameters for the 
CNGS baseline, with $E_\nu <2.5$~GeV. The $^\dagger$ cases correspond to
the {\it present CNGS design} for target, acceptance and focusing system.}
\label{tab:CNGS}
\end{table}


\section{The intrinsic $\nue$ background}
\label{sec:back}
As well known, the intrinsic
electron (anti)neutrinos in the beam are produced either in the decay of muons coming
from kaons or pions via
the chain $$\pi^+/K^+ \rightarrow \mu^+ + \nu_\mu \ , \  \mu^+ \rightarrow \nu_e +e^+ 
+\bar\nu_\mu,$$
or directly in the three-body $K_{e3}$ kaon decays. 

When looking for $\numu\ra\nue$ oscillations, this contamination  
will eventually
be the limiting factor. It is therefore essential to understand
its level and it is also worth understanding
if the beam design can be optimized to minimize this background.

Rather than the $\nue/\numu$ ratio, we decide to consider
the ratio $\numu/\sqrt{\nue}$ in order to better estimate
the effect of the backrgound on the $\numu\ra\nue$ oscillation sensitivity.

Which source of $\nue$ is relevant to our study?

Due to the large difference between the $\pi$ and $\mu$ decay lengths, the 
electron neutrino background depends on the length of the decay tunnel
$l$. This dependence should be more evident for low beam energies. 

\begin{figure}[p]
\centering
\epsfig{file=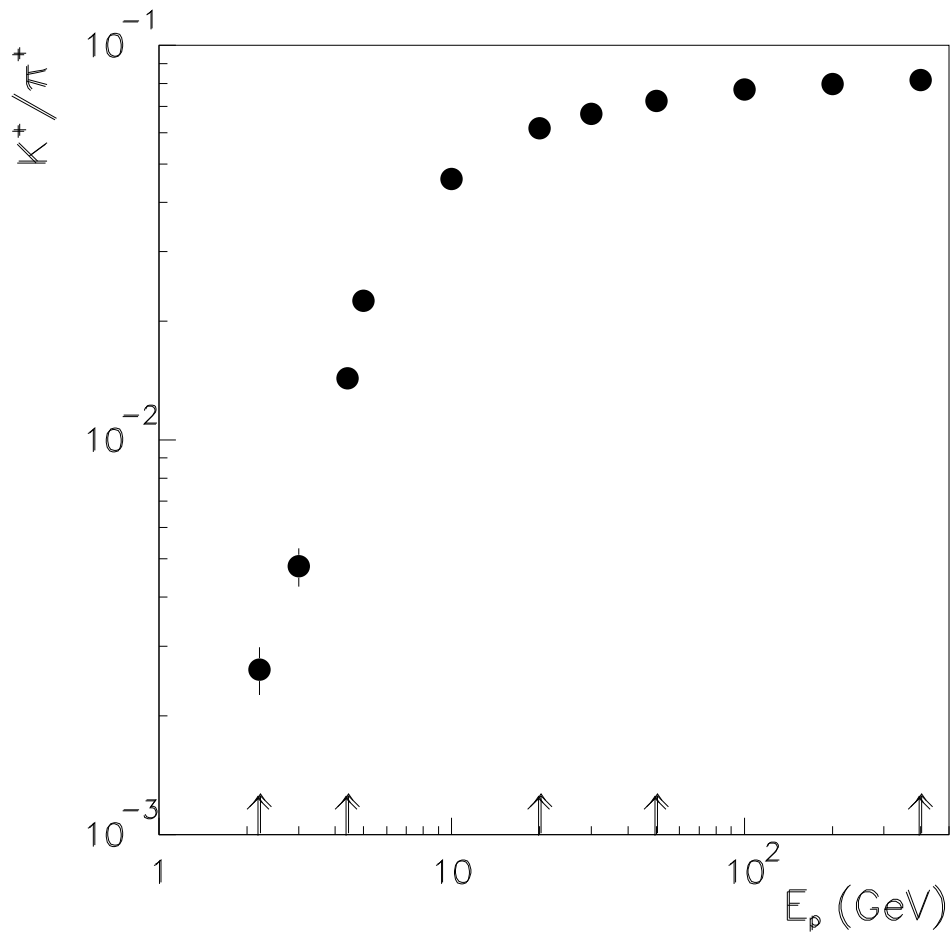,width=15cm}
\caption{Expected ratio of charged $K^+$ to $\pi^+$ production as a function of the primary proton
energy, for a 1m long, 2 mm radius graphite target. The arrows indicate proton energies
of 2.2, 4.4, 20, 50 and 400~GeV.}
\label{fig:prodk}
\end{figure}

Kaon production strongly depends on the proton energy, as
shown in Figure~\ref{fig:prodk}, where a threshold effect is
clearly visible. For the lowest proton energies, 2.2 and 4.4~GeV,
the fraction of kaon relative to pions is a few per mille. Above 20~GeV,
it is on the order of a little less than 10\%.

At low $\nu_e$ energies, the production is shared by 
kaon and muon decays,
while  kaon decays alone are responsible for the high energy tail. 
For low proton beam energies, kaon production is much lower 
and practically all the intrinsic electron contamination comes from 
muon decays.

A first order estimate of the effect of the decay tunnel length $l$
on the muon-induced background  
can be derived assuming 
forward decay at each step, and counting for each neutrino energy 
 the fraction of parent particles that decay within a path $l$.  
The fraction $D_\pi$ of $\pi$ decayed after a length $l$ is simply given by 
\begin{equation}
D_\pi=(1-e^{-\frac{l}{\lambda_\pi}})
\end{equation}
Muons have to be generated by a pion first, thus
\begin{eqnarray}
D_\mu &=&\int_0^l {\lambda_\pi^\prime e^{-\frac{y}{\lambda^\prime_\pi}} \cdot
\left ( 1-e^{-\frac{l-y}{\lambda_\mu}} \right ) dy } \nonumber \\
&=& 1 - \frac{1}{\lambda_\mu-\lambda_\pi^\prime }
\left ( \lambda_\mu e^{-\frac{l}{\lambda_\mu}} - 
\lambda_\pi^\prime e^{-\frac{l}{\lambda^\prime_\pi} }  \right )
\end{eqnarray}

The decay lengths  $\lambda^\prime_\pi, \lambda_\mu$
depend on the energy of the meson. We can fix a neutrino energy $E_\nu$, the same for
$\nu_\mu$ and $\nu_e$. 
For $\nu_\mu$ production,  we can then assume $p_\pi=E_\nu/0.43$. For the 
$\nu_e$ produced in the decay of a muon, there is no fixed relation, but we can take 
the average of the $\nu_e$ energy in the muon rest frame, getting
 $p_\mu \approx E_\nu/0.6$. The grand-parent pion had therefore
a momentum equal to
$p^\prime_\pi \approx  p_\mu \approx E_\nu/0.6$.

With these approximations, $D_\pi$ and $D_\mu$ 
can be expressed  as a function of 
$l/E_\nu$.
 In these units, one $\pi$ decay length
corresponds to $l/E_\nu =130$~m/GeV.

It is a priori obvious that
short decay tunnels reduce the
relative probability of muon production and decay. However,  the pion
decay yield is also affected. 

\begin{figure}[p]
\centering
\epsfig{file=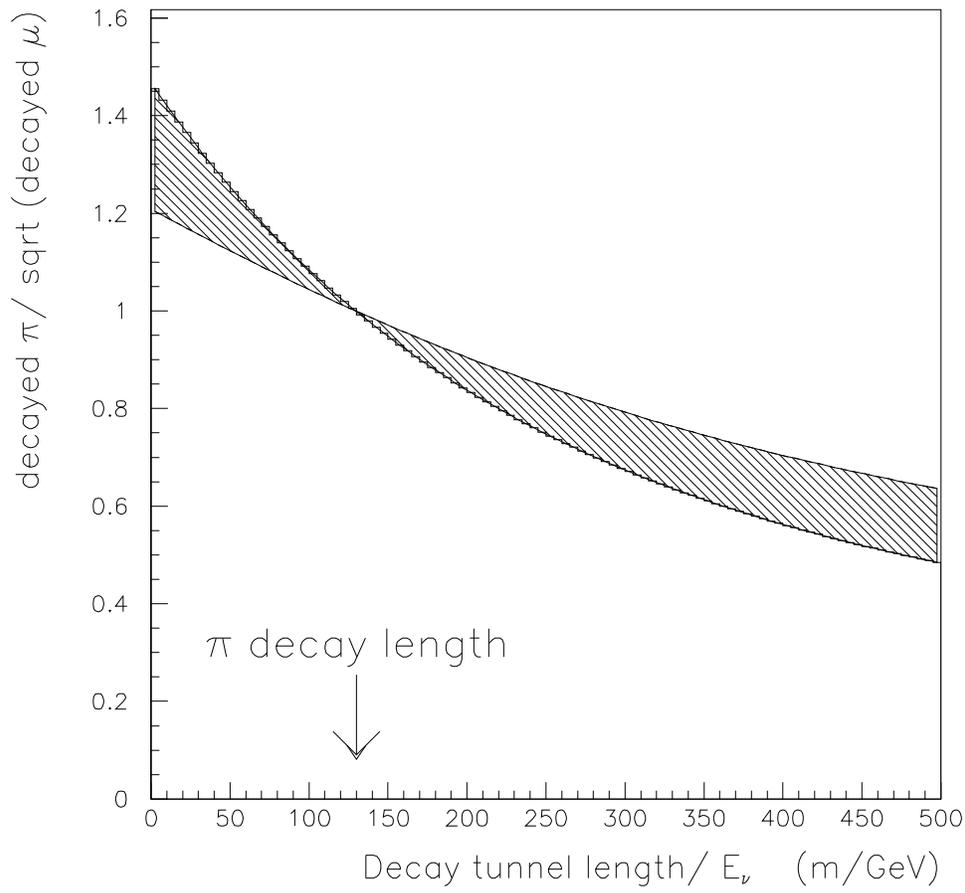,width=15cm}
\caption{
Ratio of pion decay over the square root of muon decay as a function
of decay tunnel length, normalized to the value at one pion decay length.}
\label{fig:dcytun}
\end{figure}

To study the background from muons, 
we consider the statistically significant ratio 
$\pi decays/\sqrt{\mu decays}$ as the correct estimator
for $\numu/\sqrt{\nue}$ as a function of the decay tunnel length.
This ratio is shown in Figure~\ref{fig:dcytun}, where  
the assumption of a fixed $\nu_e$ 
energy in $\mu$ decay has been relieved: 
the hatched band
corresponds to 20\%-80\% of the maximum $\nu_e$ energy in the $\mu$ rest frame.

We find that the ratio  does not show dramatic variations
between 0 and 4 pion decay lengths. We therefore conclude that {\it not much
is to be gained by reducing the length of the decay tunnel}.

We have verified these results directly by the full simulation of the neutrino
beams for various decay tunnel lengths. Answers are reported in
Table~\ref{tab:nueback} for various proton energies and decay tunnel
lengths. The 6th column shows the expected $\nue$
contamination  relative to the $\numu$ and the last column lists the statistically relevant
ratio $\numu/\sqrt{\nue}$. For the 2.2~GeV proton energy, the ratio
$\nue/\numu$ varies from 0.3\% for $l=20~m$ up to 1.7\% for $l=150~m$,
but this happens at a high cost of genuine $\numu$'s. The statistically
relevant ratio $\numu/\sqrt{\nue}$ varies from 
0.67 for $l=20~m$ down to 0.47 for $l=150~m$. This is a modest loss. We also
stress that the naive $\sqrt{N}$ scaling is not adequate for an appearance
experiment where we are looking for few events and hence we conclude
that genuine $\numu$ rate is more important than a slightly better
$\nue/\numu$ ratio.

Remains the issue of the proton energy. Naively, one would expect
that the higher the proton energy, the higher is the background.
We find however that the intrinsic electron neutrino background
does {\it not} strongly  depend on the proton energy. 

\begin{figure}[p]
\centering
\epsfig{file=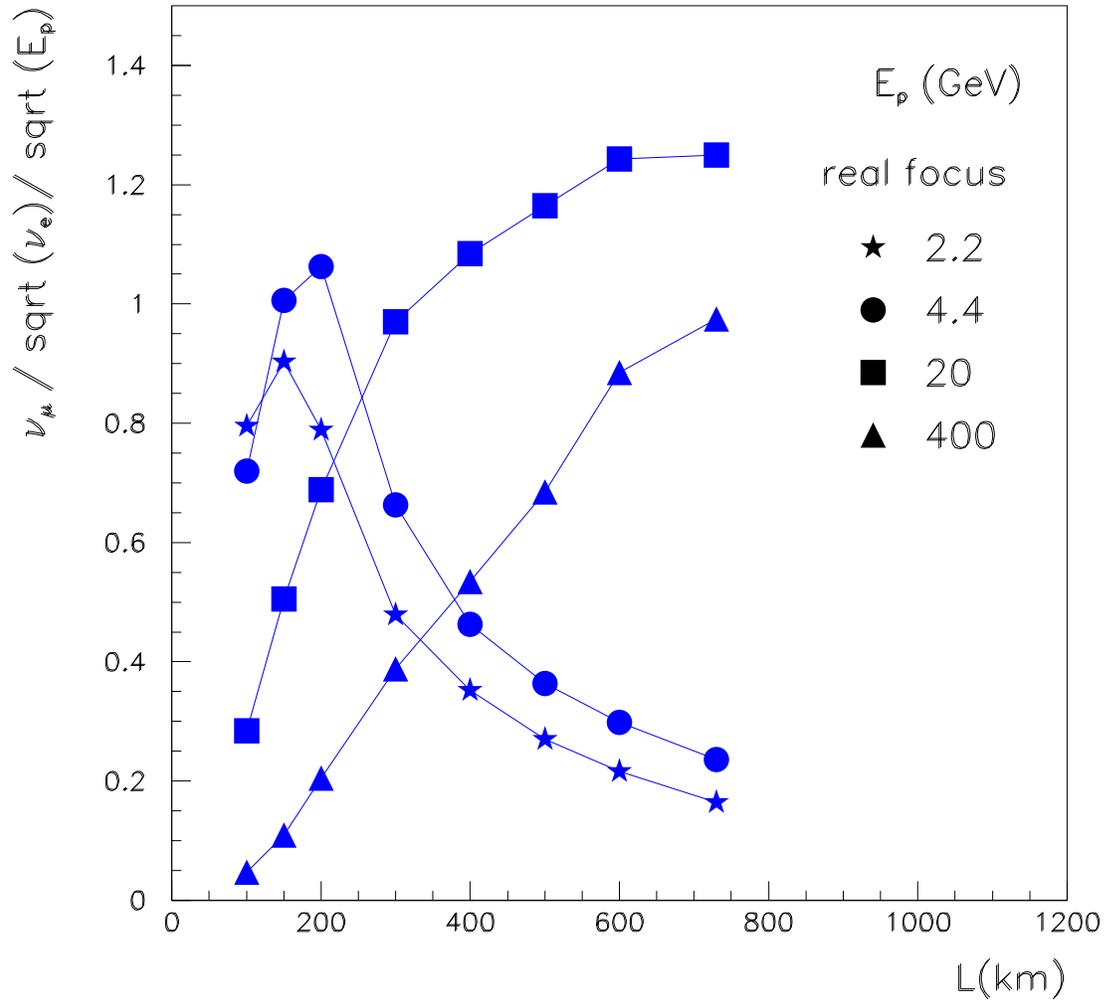,width=17cm}
\caption{Rescaled ratio of muon neutrino flux over the square root of the
electron neutrino flux divided by the square root of the proton energy (see text)
$(\numu/\sqrt{\nue})/\sqrt{E_p}$ as a function of the baseline $L$.}
\label{fig:sovbgs}
\end{figure}

This was verified directly for various proton energies and baselines.
Results of the calculation are shown in Table~\ref{tab:relat}, all normalized
to $10^{19}$ pots.
We observe that (1) for the shortest baselines, the ratio $\nue/\numu$
is increasing dramatically with proton energy. Accordingly, (2) the ratio
$\numu/\sqrt{\nue}$ decreases. However, we must rescale this ratio
to take into account the approximate scaling of the number of events with the
proton energy. Since we expect
\begin{equation}
\frac{\numu}{\sqrt{\nue}}\propto \frac{E_p}{\sqrt{E_p}} \propto \sqrt{E_p} 
\end{equation}
we consider the rescaled ratio $\numu/\sqrt{\nue\cdot E_p}$ in the last
columns of the Table~\ref{tab:relat}. These are also plotted
in Figure~\ref{fig:sovbgs} as a function of the baseline $L$.
Numerically, we find that {\it the rescaled ratios $\numu/\sqrt{\nue\cdot E_p}$
are almost the identical at the optimal baselines of each proton energy}, so
not much is too be gained by varying the proton energy.

Summarizing, we find that the electron neutrino
background in the relevant region
is not dependent on the proton energy
and only determined by the decay tunnel length, but
that its optimisation is very limited.
For the maximum neutrino muon flux, it is at the
level of the 1\% for any of the considered setups.

\begin{table}[htb]
\begin{center}
\begin{tabular}{|c|c|c|c|c|c|c|}
\hline
&&&\multicolumn{2}{c|}{ }&&\\
$E_p$&Focus&Decay tunnel  
&$\nu_\mu$ &$\nu_ e$ & $\nu_e / \nu_\mu$ &$\nu_\mu / \sqrt{\nu_e}$  \\  
GeV&& length, m&
\multicolumn{2}{c|}{ev/kton/$10^{19}$ p.o.t.}& & \\
& & & \multicolumn{2}{c|}{L=730~km} & & \\
\hline
 2.2& p.f.&150	
	&$ 3.9\times 10^{-3}$
	&$ 6.7\times 10^{-5}$
&1.7\% &0.47\\
 2.2& p.f.&50	
	&$ 2.5\times 10^{-3}$
	&$ 2.1\times 10^{-5}$
&0.84\% &0.55\\
 2.2& p.f.&20	
	&$ 1.2\times 10^{-3}$
	&$ 3.2\times 10^{-6}$
&0.27\%  & 0.67\\
 2.2& coil&150	
	&$ 1.5\times 10^{-3}$
	&$ 3.1\times 10^{-5}$
&2.1\% &0.3\\
\hline
4.4 & p.f.&150	
	&$ 2.6\times 10^{-2}$
	&$ 3.9\times 10^{-4}$
&1.5\%&1.3\\
4.4 & coil&150
	&$ 4.6\times 10^{-3}$
	&$ 7.6\times 10^{-5}$
&1.6\% &0.53\\
\hline
20 & p.f.&150
	&$ 0.56$
	&$ 4.1\times 10^{-3}$
&0.7\% &8.7\\
20 & p.f.&350
	&$ 0.9$
	&$ 0.011$
&1.3\% &8.6\\
20 &horn&350
	&$ 0.36$
	&$ 4.2\times 10^{-3}$
&1.2\% & 5.6\\
\hline
400 & p.f  &350
	&$ 9.0$
	&$ 0.12$
&1.3\% &26.0\\
400 & horn &350
	&$ 4.5$
	&$ 4.2\times 10^{-2}$
&0.9\% &22.0 \\
\hline
\hline
\end{tabular}
\end{center}
\caption{Electron neutrino intrinsic background within $E_\nu <2.5$~GeV
for various proton energies and beam optics configurations. For ease
of comparison, rates are
normalized to a baseline L=730~km.}
\label{tab:nueback}
\end{table}

\begin{table}[htb]
  \begin{center}
    \begin{tabular}{|l|c|c|c||c|c|c||c|c|c||c|c|c|}
\hline
\hline
L&\multicolumn{3}{c|}{2.2 GeV}&\multicolumn{3}{c|}{4.4 GeV}
&\multicolumn{3}{c|}{20 GeV}
&\multicolumn{3}{c|}{400 GeV}\\
Km&\multicolumn{3}{c|}{coil focus}&\multicolumn{3}{c|}{coil focus}
&\multicolumn{3}{c|}{horn focus}
&\multicolumn{3}{c|}{horn focus}\\
 & {\large $\frac{\nu_e}{\nu_\mu}$}(\%) 
   &{\large $\frac{\nu_\mu}{\sqrt{\nu_e}}$}
   &{\large $\frac{\nu_\mu}{\sqrt{\nu_e\cdot E_p}}$} 
 & {\large $\frac{\nu_e}{\nu_\mu}$}(\%) 
   &{\large $\frac{\nu_\mu}{\sqrt{\nu_e}}$}
   &{\large $\frac{\nu_\mu}{\sqrt{\nu_e\cdot E_p}}$} 
 & {\large $\frac{\nu_e}{\nu_\mu}$}(\%)
   &{\large $\frac{\nu_\mu}{\sqrt{\nu_e}}$}
   &{\large $\frac{\nu_\mu}{\sqrt{\nu_e\cdot E_p}}$} 
 & {\large $\frac{\nu_e}{\nu_\mu}$}(\%) 
   &{\large $\frac{\nu_\mu}{\sqrt{\nu_e}}$}
   &{\large $\frac{\nu_\mu}{\sqrt{\nu_e\cdot E_p}}$} 
 \\
\hline
 100
         &    1.6  &   1.2  &   0.79
         &    1.6  &   1.5  &   0.72
         &    5.9  &   1.3  &   0.28
         &   25.7  &   0.9  &   0.05
\\
 150
         &    1.4  &   1.3  &   0.91
         &    1.2  &   2.1  &   1.00
         &    4.1  &   2.3  &   0.51
         &   13.5  &   2.2  &   0.11
\\
 200
         &    1.4  &   1.2  &   0.79
         &    1.1  &   2.2  &   1.06
         &    3.2  &   3.1  &   0.69
         &    7.6  &   4.1  &   0.20
\\
 300
         &    1.8  &   0.7  &   0.48
         &    1.4  &   1.4  &   0.66
         &    2.0  &   4.3  &   0.97
         &    3.3  &   7.8  &   0.39
\\
 400
         &    1.8  &   0.5  &   0.35
         &    1.6  &   1.0  &   0.46
         &    1.6  &   4.9  &   1.08
         &    2.1  &  10.7  &   0.54
\\
 500
         &    1.9  &   0.4  &   0.27
         &    1.6  &   0.8  &   0.36
         &    1.4  &   5.2  &   1.16
         &    1.6  &  13.7  &   0.68
\\
 600
         &    2.0  &   0.3  &   0.22
         &    1.6  &   0.6  &   0.30
         &    1.2  &   5.6  &   1.24
         &    1.2  &  17.7  &   0.89
\\
 730
         &    2.1  &   0.2  &   0.16
         &    1.7  &   0.5  &   0.24
         &    1.1  &   5.6  &   1.25
         &    1.0  &  19.5  &   0.97
\\
\hline
   \end{tabular}
  \end{center}
\caption{ Relationships between muon neutrino CC events  and 
electron neutrino CC events,  for 10$^{19}$ pots. 
Event spectra have been integrated over the energy range of
interest at each baseline $L$.}
\label{tab:relat}
\end{table}


\section{Summary and Conclusions}

 In this document, we have performed a two-dimensional  scan, varying   the 
beam energy and baseline parameters to optimize  the conditions for   
 the investigation of $\theta_{13}$ driven neutrino oscillations in the whole Superkamiokande
allowed $\Delta m^2$ range. We find that:

\begin{itemize}
\item The optimal baselines  for $\theta_{13}$ searches are in the range 
100-700~km for proton energies varying from 2.2 to 400~GeV.

\item The needed beam intensity scales approximately with the inverse of
the beam energy.  
  
\item In terms of proton economics, the optimum beam energy is around 20~GeV, 
but lower beam energies are appealing for the shortest baselines.

\end{itemize}

Realistic focusing system for low and medium baselines 
 have also been studied. In this case:

\begin{itemize}
\item Focusing efficiencies of 30-50\% can be achieved in the energy range of
interest.
\end{itemize}

The whole procedure is detector and accelerator independent. Nevertheless, 
to give  a first estimate of the needed beam intensities, the ICARUS
detector has been assumed as a reference.

We can draw the following observations:
\begin{itemize}
\item a 2.2 GeV or 4.4 GeV high-intensity proton machine (i.e. \`a la CERN-SPL) is well matched
to a baseline in the range 100-300~km. It is not matched to a baseline of 730~km (i.e. CERN-LNGS).
\item a 20 GeV machine is best matched to a baseline of 730~km (i.e. CERN-LNGS).
However, an integrated intensity in the range of $10^{23}$ pots are required, 
which is about two orders of magnitude
higher than the intensity deliverable by the current CERN-PS in a reasonable amount of time.
\item a 400~GeV energy is reasonably matched to a baseline of 730~km (i.e. CERN-LNGS). For
the 400~GeV, the required intensity is in the range of $10^{22}$ pots, which is about one order of magnitude
higher than the intensity deliverable by the current CERN-SPS in a reasonable amount of time.
\end{itemize}

Finally, we stress that the present study is essentially a theoretical one.
All the ``real'' work has still to be accomplished in order for one of these
options to become reality.

%
%


\begin{figure}[p]
\centering
\epsfig{file=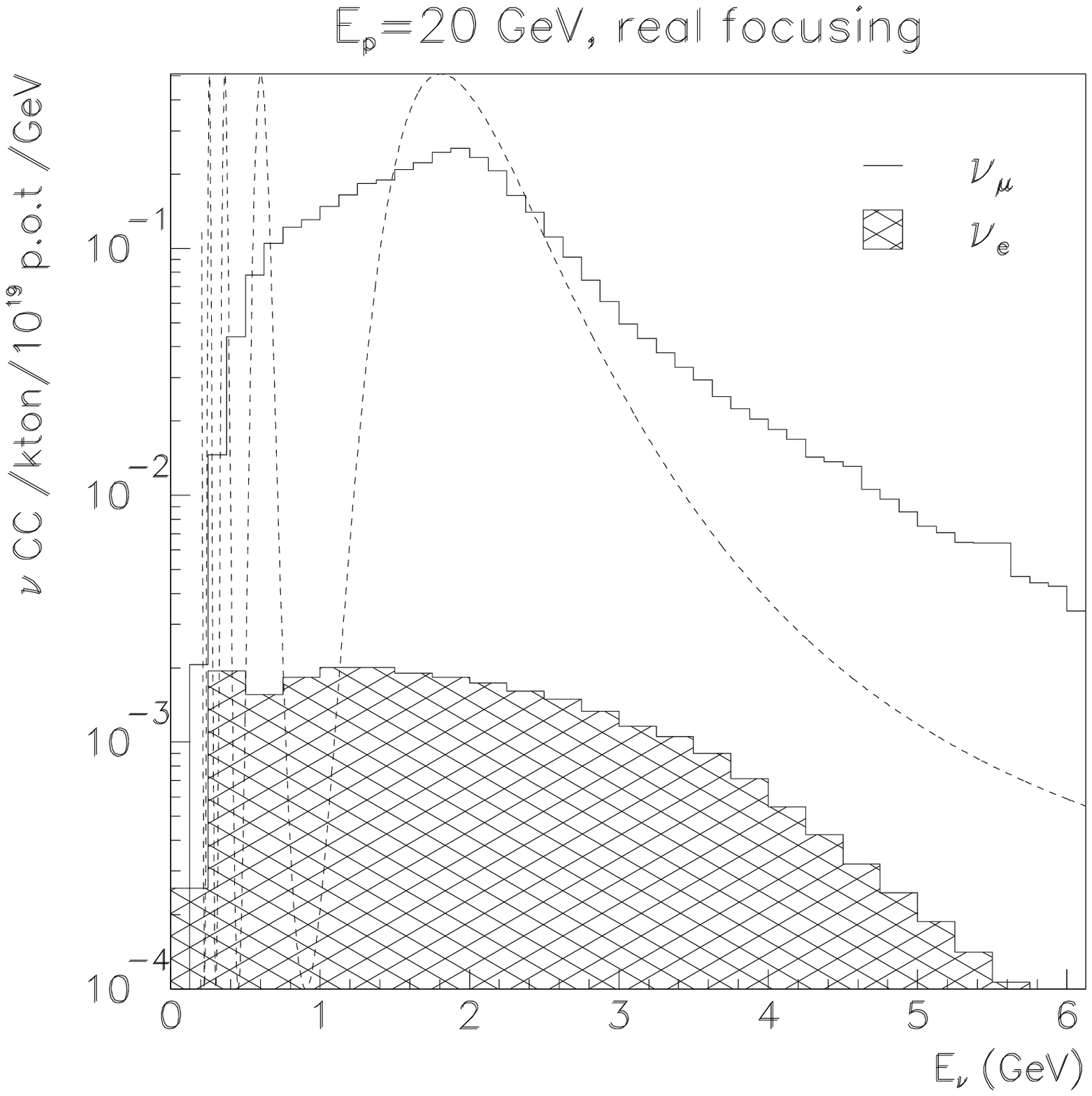,width=15cm}
\caption{ 
Charged current event rate at 730 km as a function of the neutrino energy
computed with (real) horn focusing for 20~GeV proton energy. The dotted line corresponds
to the oscillation probability (arb. norm.) for a $\Delta m^2=3\times 10^{-3}\rm\ eV^2$.}
\label{fig:plotenesplps_e}
\end{figure}

\end{document}